\documentstyle[12pt]{article}
\def\sq{{\vbox {\hrule height 0.6pt\hbox{\vrule width 0.6pt\hskip 3pt
   \vbox{\vskip 6pt}\hskip 3pt \vrule width 0.6pt}\hrule height 0.6pt}}}
\newcommand{\roughly}[1]{\raise.3ex\hbox{$#1$\kern-.75em
\lower1ex\hbox{$\sim$}}}

\def\tf#1#2{{\textstyle{#1 \over #2}}}
\def\df#1#2{{\displaystyle{#1 \over #2}}}

\def\overleftrightarrow#1{\vbox{\ialign{##\crcr
     $\leftrightarrow$\crcr\noalign{\kern-0pt\nointerlineskip}
     $\hfil\displaystyle{#1}\hfil$\crcr}}}

\def\lsim{\mathrel{\mathstrut\smash{\ooalign{\raise2.5pt\hbox{$<$}\cr\lower2.5pt\hbox{$\sim$}}}}}
\def\gsim{\mathrel{\mathstrut\smash{\ooalign{\raise2.5pt\hbox{$>$}\cr\lower2.5pt\hbox{$\sim$}}}}}

\def\sqr#1#2{{\vcenter{\vbox{\hrule height.#2pt
         \hbox{\vrule width.#2pt height#1pt \kern#1pt
            \vrule width.#2pt}
         \hrule height.#2pt}}}}
\def\square{\mathop{\mathchoice\sqr56\sqr56\sqr{3.75}4\sqr34\,}\nolimits}

\def\Re{{\rm Re}\,}
\def\Im{{\rm Im}\,}

\def\beq{\begin{equation}}
\def\eeq{\end{equation}}
\begin{document}
\baselineskip=15.5pt
\pagestyle{plain}
\setcounter{page}{1}
\begin{titlepage}
\bigskip
\rightline{NSF-ITP-02-71}
\rightline{SU-ITP-02/27}
\rightline{hep-th/0208123}
\bigskip\bigskip\bigskip\bigskip
\centerline{\Large \bf Scales and hierarchies in warped }
\smallskip
\centerline{\Large \bf compactifications and brane worlds}
\bigskip\bigskip
\bigskip\bigskip

\centerline{\large  Oliver DeWolfe$^a$\footnote{\tt odewolfe@itp.ucsb.edu} and Steven B.~Giddings$^{b,c}$\footnote{\tt giddings@physics.ucsb.edu}}
\bigskip
\centerline{\em $^a$Kavli Institute for Theoretical Physics and $^b$Department of Physics\footnote{Current address.},}
\centerline{\em UCSB, Santa Barbara, CA 93106}
\medskip
\centerline{\em $^c$Department of Physics and SLAC, Stanford University, Stanford, CA 94305/94309}
\medskip
\bigskip\bigskip


\begin{abstract}
Warped compactifications with branes provide a new approach to the
hierarchy problem and generate a diversity of four-dimensional thresholds.
We investigate the relationships between these scales, which fall into two
classes.  Geometrical scales, such as thresholds for Kaluza-Klein, excited
string, and black hole production, are generically determined soley by the
spacetime geometry.  Dynamical scales, notably the scale of supersymmetry
breaking and moduli masses, depend on other details of the model.  We
illustrate these relationships in a class of solutions of type IIB string
theory with imaginary self-dual fluxes.  After identifying the geometrical
scales and the resulting hierarchy, we determine the gravitino and moduli
masses through explicit dimensional reduction, and estimate their value to
be near the four-dimensional Planck scale.  In the process we obtain
expressions for the superpotential and K\"ahler potential, including the
effects of warping.  We identify matter living on certain branes to be
effectively {\em sequestered} from the supersymmetry breaking fluxes:
specifically, such ``visible sector'' fields receive no tree-level masses
from the supersymmetry breaking.  However, loop corrections are expected
to generate masses, at the phenomenologically viable TeV scale.

\medskip
\noindent
\end{abstract}
\end{titlepage}

\section{Introduction}
\label{IntroSec}

Recent years have opened up a new universe of string
compactifications.  Much of the work done on string phenomenology
after the ``first superstring revolution'' of 1984 had focused on
traditional Kaluza-Klein compactifications of string theory to four
dimensions.  However, we now 
see a great range of extensions of this picture: one may first of all
consider more generally {\it warped compactifications}, and secondly
one may have {\it brane world} scenarios in which branes -- wrapped or
otherwise -- are present.
This leads to a wide new spectrum of possibilities for reproducing
four-dimensional Poincar\'e invariant physics from higher-dimensional
string/M-theory.  Particularly interesting are the resulting
geometrical/dynamical mechanisms that allow the string scale to be
many orders of magnitude lower than the traditional value $\sim
10^{19}$ GeV -- and perhaps even as low as ${\cal O}(1\, {\rm TeV})$,
providing a completely new potential resolution of the hierarchy
problem.  We still seem only to have scratched the surface in
exploring this new universe.

The added complexities of these models imply the possibility of various new
phenomena taking place at differing scales.  In the case where some of
these thresholds are lowered to ${\cal O}(1\, {\rm TeV})$ -- or even lower
-- clearly it is especially interesting to understand what they are, and
how they are related to the geometry and fields on the internal manifold.
The diversity of possible scales include the natural scale for scalar
masses, the apparent and fundamental Planck scales, thresholds for
production of Kaluza-Klein states, excited string states, or microscopic
black holes, and the supersymmetry breaking scale.  We will discuss the
emergence of these in general warped compactifications/brane worlds that
occur in string/M-theory.

Models exhibiting these phenomena include the large extra dimensions
scenario of \cite{Arkani-Hamed:1998rs} and the warped model of
\cite{Randall:1999ee}.  Although inspired by stringy developments, the
original proposals were not directly related to an underlying microscopic
theory, but were solutions of effective theories capturing essential ideas.
Large extra dimensions were subsequently discussed in the context of string
theory in \cite{Antoniadis:1998ig}, and more complete embeddings of warped
scenarios have emerged.  As a specific example, \cite{Giddings:2001yu}
provides a string solution that geometrically realizes a hierarchically low
fundamental string scale via warping, along the lines of
\cite{Randall:1999ee}.  A warped geometry is created within a Calabi-Yau
threefold by fluxes in the spirit of
\cite{Verlinde:1999fy,Chan:2000ms,Klebanov:2000hb}, with a throat that
comes to a smooth end playing the role of an infrared brane, while the
Calabi-Yau manifold itself plays the role of an ultraviolet brane by
terminating the throat at the top.  Since the total space is compact, this
picture bears similarities to both \cite{Arkani-Hamed:1998rs} and
\cite{Randall:1999ee}.  In these theories, the fluxes have the additional
benefit of freezing many geometric moduli of the Calabi-Yau background, as
well as the dilaton (see also \cite{Dasgupta:1999ss}).  The ``Gukov-Vafa-Witten'' (GVW) superpotential
\cite{Gukov:1999ya} that freezes these moduli also can break supersymmetry
spontaneously.

In this paper we study the relationship of the various thresholds of
physical phenomena in a warped/brane world compactification, both to
each other, and to properties of the underlying geometry.  Several of
these scales rely only on simple properties of the geometry, and very
general statements can be made.  We refer to these as {\it geometrical
scales.}  Some of the scales, particularly involving supersymmetry breaking,
are less generic and more model dependent; we refer to these as {\it
dynamical scales}. We will illustrate some of this model dependence in
the context of the model of \cite{Giddings:2001yu} and the GVW
superpotential.

In outline, we begin with a brief general discussion of warped
compactifications and brane worlds.  We follow this by discussing the
general relationships between geometrical scales: the fundamental and
apparent four-dimensional Planck scales, the string scale, and the
typical mass-scales for brane matter.  This is essentially a simple
extension of known results.  We then discuss the more model-dependent
(but still geometrical) question of the thresholds for Kaluza-Klein
modes.  We next turn to dynamical scales, particularly the
supersymmetry breaking scale.  Here the observed scales depend
sensitively both on the {\it form} of supersymmetry breaking
(e.g. gravity-mediated from the moduli sector or a hidden brane
sector, or gauge-mediated from extended gauge dynamics on the branes),
and on the warping in the region where it is localized.

We then give an extensive illustration of our comments in the context of
the compactifications of \cite{Giddings:2001yu}.  With a moderate choice of
discrete fluxes, these solutions generate a hierarchy between the weak and
Planck scales, while at the same time breaking supersymmetry and fixing
many of the problematic moduli familiar from traditional Calabi-Yau
compactifications.  After outlining properties of these solutions, we
derive expressions for the gravitino mass and for the potential for moduli.
While not essential for the derivation, these can be thought of as arising
from a four-dimensional effective supergravity action, and we exhibit the
corresponding K\"ahler and superpotentials, explicitly including the
effects of warping.  Generically the gravitino and moduli masses are
estimated to be large, of order $M_4\sim 10^{19}$ GeV, an apparent
phenomenological disaster.  However, as a result of no-scale structure,
tree level masses for scalars living on an IR brane vanish.  Moreover,
fermion masses also vanish at tree level \cite{Grana:2002tu}, producing a
close analog of the {\it sequestered} scenarios of \cite{Randall:1998uk}.
To our knowledge this is the first realization of sequestering in a string
theory background.  The sequestered form persists even incorporating brane
backreaction, though it may not survive $\alpha'$ corrections.  These
``visible sector'' masses receive contributions from loops; the warped
structure of the solution indicates that these corrections should be of
order ${\cal O}$(TeV) for solutions where the hierarchy is indeed generated
by warping.  Section~\ref{StringSec} is rather long and technical, but the
reader interested in a brief overview is directed to a summary in
subsection~\ref{SummarySec}.

\section{Warped geometries, brane worlds, and the hierarchy}
\label{WarpedSec}

In traditional Kaluza-Klein compactifications, the extra dimensions
$y^m$, $m=1,\cdots, D-4$ of $D$-dimensional space time (or more
generally, in string theory the extra-dimensional conformal field
theory) are taken to form a direct product geometry with the visible
dimensions $x^\mu$, $\mu=0,1,2,3$:
\begin{equation}
ds^2= \eta_{\mu\nu}dx^\mu dx^\nu + g_{mn}(y)\, dy^m dy^n\,,
\end{equation}
where $g_{mn}$ is the metric in the extra dimensions.
However, we have increasingly realized the potential importance of
compactifications in which this geometry is extended to the most
general 4d Poincar\'e-invariant form:
\begin{equation}
ds^2= e^{2A(y)}\eta_{\mu\nu}dx^\mu dx^\nu + g_{mn}(y)\, dy^m dy^n\,. 
\label{warpc}
\end{equation}
Such a compactification is known as a {\it warped compactification},
and the function $e^{2A}$ as a {\it warp factor}
\cite{Cremmer:1979up, deWit:1984va, vanNieuwenhuizen:ri,Duff:hr}.

A second important extension, following from the ``second superstring
revolution,'' is the inclusion of branes.  In order to preserve 4d
Poincar\'e invariance, these should be fully extended over the dimensions of
observed four-dimensional spacetime.  Their configuration in the extra
dimensions is more flexible.  Simplest is the case of D3 branes, which then
are pointlike in the extra dimensions.  But more generally, the compact
geometry can have non-trivial closed cycles on which some of the dimensions
of a D$p$-brane, with $p>3$, can wrap.  

Within the context of string theory, there are also higher-form 
antisymmetric tensor
fields that can acquire vevs in the compact directions, without spoiling
Poincar\'e invariance.  

D-branes, fluxes, and warping are of course in general related, since
D-branes serve as sources for fluxes, and both D-branes and fluxes may
serve as sources of non-trivial warp factors.  It is also possible for
D-branes and fluxes to transmute into one another.

Our interest is in string/M theory propagating on the spacetime
(\ref{warpc}).  As long as geometrical features are larger than the
fundamental Planck length, the dynamics is well-described in terms of a
low-energy effective action of the form
\begin{equation}
S={M_D^{D-2}\over (2\pi)^{D-4}} \int d^D x \sqrt{-g} {1 \over 2} {\cal R}+ \int d^D x
\sqrt{-g} {\cal L}\,,
\end{equation}
where $M_D$ is the fundamental Planck mass (in the phenomenologically
useful conventions of \cite{Abe:2001wn}), ${\cal R}$ is the Ricci
scalar, and ${\cal L}$ is the Lagrangian for other fields and sources,
including matter, fluxes, and branes.  

We would like to determine the parameters that govern four-dimensional
phenomenology, in terms of the parameters of the underlying
fundamental theory.  In the example of a string compactification of
the type II string, for which $D=10$, the string frame Lagrangian
takes the form
\begin{equation}
S\propto{M_s^8}
\int d^{10} x \sqrt{-g} e^{-2\phi} \, {\cal R} + \cdots\,, \label{action}
\end{equation}
up to a numerical constant,
where $\phi$ is the dilaton and the string coupling is $g_s=e^{\langle
\phi\rangle }$.  The relation between the fundamental string scale and the
Planck scale immediately follows:
\begin{equation}
M_{10} = g_s^{-{1\over 4}} M_s\,. \label{mstring}
\end{equation}
The relation between the fundamental and apparent 4-dimensional Planck
scales is nearly as simple.  Indeed, replace the metric (\ref{warpc}) by
one including 4d fluctuations
\begin{equation}
ds^2= e^{2A(y)}g_{\mu\nu}dx^\mu dx^\nu + g_{mn}(y)\, dy^m dy^n\,, 
\end{equation}
and substitute into the action (\ref{action}).  We find that fluctuations
of the 4d metric about internal geometries obeying the equations of motion
are governed by an effective action
\begin{equation}
S_4= {M_4^2 \over 2} \int d^4 x \sqrt{-g_4(x)} {\cal R}_4 \,,
\end{equation}
with the four- and D-dimensional Planck masses related by
\begin{equation}
{M_4^2\over M_D^2} = \left({M_D \over 2 \pi} \right)^{D-4} \int d^{D-4}y \sqrt{g_{D-4}} e^{2A}\equiv \left({M_D \over 2 \pi} \right)^{D-4}
V_w\,.\label{mplanck}
\end{equation}
This equation defines the ``warped volume'' $V_w$.

Next consider mass scales for matter fields.  In particular, if fermion
masses are generated by a Higgs scalar $H$, in the absence of a protection
mechanism, radiative corrections are expected to generate scalar masses
$M_H$ of order the cutoff, which here is expected to be ${\cal O}(M_D)$,
in the Lagrangian ${\cal L}$.  In the general brane world scenario, where
fermion and Higgs fields propagate on a ``standard model'' 
$p$-brane with coordinates $z$, 
this results in a
contribution to the action of the form
\begin{equation}
S_H= - {1\over 2}\int d^4 x \int d^{p-4} z \sqrt{g_{\rm brane}}
\left[e^{2A}(\nabla_\mu H)^2 +M_H^2 e^{4A}H^2\right]\,, \label{Higgsact} 
\end{equation}
where $g_{\rm brane}$ is the induced metric on the brane.  
From this we
find that the Higgs mass scale is given in terms of averages of the
warp-factor over the standard-model brane, by
\begin{equation}
M_0^2= \frac{\int d^{p-4} z \sqrt{g_{\rm brane}} e^{4A}}{\int d^{p-4}
z \sqrt{g_{\rm brane}} e^{2A}} M_H^2 \sim e^{2A_{\rm SM}} M_D^2\
\label{scalmass}
\end{equation}
where we denote the average of the warp factor on the standard model (SM)
or visible brane by
\begin{equation}
e^{2A_{\rm SM}} = \int d^{p-4} z \sqrt{g_{\rm brane}}  e^{2A}\,.
\end{equation}
This makes it clear that fields localized in regions where $e^A\ll 1$ have
their masses suppressed relative to the fundamental scale $M_D$; natural TeV
scale masses can be generated by the warp factor.

An alternative viewpoint of this mechanism comes from using the Weyl
symmetry of the
actions (\ref{action}), (\ref{Higgsact}).  Define the new variables
\begin{equation}
g = \lambda^2 \bar{g}\,;\ (M_D,M_0) = (\bar{M}_D,\bar{M}_0)/\lambda\,; H = \bar{H}/\lambda \,,
\end{equation}
with corresponding scalings for other fields and dimensionful
parameters.  This choice of scale may be used to set the average
\begin{equation}
e^{2\bar{A}_{\rm SM}}=1\,. \label{smconv}
\end{equation}
In these units (barring a large ratio of the different averages that enter in
(\ref{Higgsact})), the fundamental Planck scale and the Higgs mass are both
naturally comparable, and we therefore have a choice:
\begin{enumerate} 
\item{\it Conventional Planck-scale compactification:} Take
$\bar{M}_D \sim M_4 \sim 10^{19}$ GeV, and find a mechanism, such as
supersymmetry, to suppress the Higgs mass to a far smaller scale;
\item {\it TeV-scale gravity scenario:} Take $\bar{M}_D \sim 1$ TeV,
which then requires $V_w\gg 1/{\bar{M}_D}^{D-4}$.
\end{enumerate}
From the definition (\ref{mplanck}) of the warped volume and the
convention (\ref{smconv}) we see that the latter choice results from
either large volume or a large warp factor away from the brane, or
some combination of the two; in the barred variables, these two
effects are on the same footing.

Thus to summarize, there are two possible conventions from which to
understand the physics of the hierarchy in the context of a TeV-scale
gravity model.  In the first, the fundamental scale is $M_D\sim 10^{19}$
GeV, and scalar masses are suppressed to a TeV.  The second corresponds to
a definition of four dimensional energy relative to an observer localized
on the brane; for such an observer, the fundamental scale is reached at
four-dimensional energies $\bar{M}_D \sim TeV$, and this is also the
natural scale for scalar masses.  The four-dimensional Planck scale $M_4$
is enhanced relative to these by the large warp factor in (\ref{mplanck}).
We will find the barred variables to be convenient for most the the
following sections, although we will revert to the unbarred variables for
the purposes of calculating masses of bulk fields in
section~\ref{StringSec}.

\section{Geometrical scales and thresholds}
\label{GeomSec}

In conventional Planck-scale compactifications, many of the new phenomena
resulting from the compactification are only accessible in the vicinity of
the four-dimensional Planck scale, $M_4\sim 10^{19}$ GeV.  One of the
reasons for the great interest in warped compactifications is the much
greater latitude in the possible scales at which observable phenomena may
occur.  Many of these scales are determined purely from the
\textit{geometry} of the warped compactification, as opposed to other
dynamical information.  We have just seen two examples:  the relationships
between the fundamental Planck scale, the apparent four-dimensional Planck
scale, and the naturalness scale for scalar masses are determined through
relations  (\ref{mplanck}) and (\ref{scalmass}) and depend only on the warp factor
and the geometry of the internal manifold.  
In this section we will extend this discussion of physical scales and the
corresponding thresholds for other physical phenomena.

\subsection{Strings and black holes}

The most exciting possibility raised by warped compactification is that, as
outlined above, the fundamental Planck scale may be much lower than the
apparent four-dimensional Planck scale.  This means that we may begin to
experimentally access the dynamics of quantum gravity much sooner than
previously anticipated.

For example, it is believed that the generic high-energy physics of gravity
is the production of black holes.  If, as in the preceding section, we work
in units where the warping is unity on the IR brane, the fundamental Planck
scale $\bar{M}_D$ may be as low as a TeV.  Of course, the fundamental Planck
scale generically represents the threshold for production of microscopic
black holes, so above this energy collisions of particles on the SM brane
can produce black
holes; this corresponding phenomenology is discussed in 
\cite{Giddings:2000ay,Giddings:2001bu,Dimopoulos:2001hw}\footnote{It has
long been believed that collisions above the Planck energy should create
black holes.  An early concrete statement is Thorne's hoop
conjecture \cite{Thorne:ji}, and such processes were further studied in
\cite{Penrose} and \cite{D'Eath:hb,D'Eath:hd,D'Eath:qu}.
Ref. \cite{Banks:1999gd} pointed out the relevance of such black hole
formation within the TeV-scale gravity models of
\cite{Arkani-Hamed:1998rs}, and discussed some 
aspects of the phenomenology.
Other aspects of black holes in these models were discussed in
\cite{Argyres:1998qn}, and their evaporation in \cite{Emparan:2000rs}.  The
experimental relevance of black hole formation in warped scenarios was
pointed out in \cite{Giddings:2000ay}.  A general argument for classical 
black hole
formation at high energies 
appears in \cite{Eardley:2002re}.  For reviews, see
\cite{Giddings:2001ih,Giddings:2002av,Giddings:2002kt}.}.

In the context of string theory, this threshold may be pushed up to make
room for an intermediate regime where string states are produced.  This
depends on the value of $g_s$.  As we see from (\ref{mstring}), at weak
coupling the threshold for string production is below the Planck energy.
At the same time, the string length exceeds the Planck length,
\begin{equation}
l_s\sim g_s^{-1/4} l_p\,.
\end{equation}
Objects smaller than this will explicitly exhibit behavior characterized by
non-local string dynamics, and
classical black holes will only begin to exist once their radii exceed this
value, at the \textit{correspondence} scale \cite{Horowitz:1996nw}
\begin{equation}
M_c\sim \frac{M_s}{g_s^2}\,.
\end{equation}
Between $M_s$ and $M_c$ we expect perturbative string states gradually
to become more strongly coupled and morph into black hole states,
perhaps with intermediate states best described as ``string
balls" \cite{Bowick:1989us,Dimopoulos:2001qe}.

So, to summarize the results of this subsection, for weakly coupled string
theory, we should start seeing perturbative string states at the threshold
$g_s^{1/4}M_{10}$; these become more strongly coupled, and evolve into the
generic gravitational physics of black holes above the threshold
$M_s/g_s^2$.  Some of the phenomenology of the initial perturbative string
regime has been discussed in \cite{Cullen:2000ef}.

\subsection{Kaluza-Klein modes}

Another generic phenomenon is production of Kaluza-Klein modes.  For
simplicity we just discuss these in the case of scalar fields, although
results for higher-spin fields should be qualitatively similar.

Specifically, consider a $D$-dimensional scalar field $\Phi$, with action
\begin{eqnarray}
S_\Phi & = & -\frac{1}{2}\int d^D x\sqrt{-g} \left[(\nabla\Phi)^2 +
M_\Phi^2 \Phi^2\right] \\ & = & -\frac{1}{2}\int d^4x\sqrt{-g_4}\int
d^6y \sqrt{g_6} e^{4A}\left[ e^{-2A} \eta^{\mu\nu} \nabla_\mu \Phi
\nabla_\nu\Phi + g^{mn} \nabla_m\Phi \nabla_n\Phi + M_\Phi^2
\Phi^2\right]\,.
\nonumber
\end{eqnarray}
This gives an equation of motion
\begin{equation}
\sq_4 \Phi + e^{-2A}\left[ \nabla^m \left(e^{4A} \nabla_m \Phi\right) -M_\Phi^2 e^{4A} \Phi \right]=0\,.
\end{equation}

Therefore masses of Kaluza-Klein states will be given by the
eigenvalues of the wave operator
\begin{equation}
e^{-2A}\left[ \nabla^m \left(e^{4A} \nabla_m Y_i(y)\right) -M_\Phi^2
e^{4A} Y_i(y)\right]= -M_i^2 Y_i(y)\,.
\end{equation}
The size of these masses for Kaluza-Klein modes localized in the
vicinity of the SM brane is generically determined by the scales on
which the 6d metric and warp factor $A$ vary.  For example, in an
unwarped compactification, the lightest scale is roughly $1/L$, where
$L$ is the size of the largest dimension.  In the case of the
model of \cite{Randall:1999ee}, the Kaluza-Klein masses are of size
$1/R$, where $R$ is the AdS radius, in other words the scale of
variation of the warp factor which in this case is just
\begin{equation}
A=-y/R\,;
\end{equation}
a similar result is found for the string solutions of
\cite{Giddings:2001yu} which have an approximately AdS region.
Either kind of geometrical scale will typically be larger than the
fundamental length scale (otherwise a geometrical description may not
apply), so the Kaluza-Klein masses will typically be below the
fundamental scale, even far below as in the extreme case of
\cite{Arkani-Hamed:1998rs}.

Of course, there may be more complicated scenarios where contributions
of the warp factor relative to that on the SM brane rescale these
masses.  Kaluza Klein modes localized in a region with warp factor $A$
will have their masses scaled by $e^A$.  For example,
ref.~\cite{Dimopoulos:2001ui} investigates scenarios with multiple
throats that are approximately anti de-Sitter; if we consider the KK
modes localized in throat $j$ in a vicinity with warp factor $A_j$,
then the corresponding masses will be renormalized by the factor
$e^{A_j}$ as seen by an observer on the SM brane.  Of course couplings
of such modes in a distinct throat to those of the visible sector are
expected to be correspondingly suppressed.

\subsection{Summary of geometric thresholds} 

To summarize the results of this section, in a TeV-scale gravity
scenario with hierarchy generated by warping, the sequence of
thresholds is as follows.  (This summary is given in the
``brane-based" conventions outlined in the preceding section.)  The
lowest energy threshold is generically that for Kaluza-Klein states,
\begin{equation}
M_{KK}\sim \frac{e^{A_{KK}} }{R}
\end{equation}
where $A_{KK}$ is the warp factor in the region where the state is
localized, and $R$ is a characteristic proper geometrical scale.  Next, in
the case of a string scenario, and at least for moderately weak string
coupling, comes the threshold for producing string states:
\begin{equation}
M_S\sim g_s^{1/4} M_{10} \,.
\end{equation}
For the possibly more realistic case of strong string coupling, this is
degenerate with the fundamental Planck scale, which as measured by
observers on the standard model brane is $\bar{M}_D \sim$ TeV; this is the
approximate threshold for producing black holes.  Scalar masses are also
naturally of this size:
\begin{equation}
M_0 \sim \bar{M}_D \,.
\end{equation}
The four-dimensional Planck scale lies far beyond, at
\begin{equation}
\frac{M_4^2}{\bar{M}_D^2} = \left({\bar{M}_D \over 2 \pi}\right)^{D-4} V_w\,.
\end{equation}

\section{Dynamical scales; supersymmetry breaking}

Certain physical thresholds are determined by more detailed dynamical
information than that contained in the metric; these are the dynamical
scales.  An obvious example is that of supersymmetry breaking mass scales:
the mass of the gravitino, and of superpartners.  
Moreover, generic Calabi-Yau compactifications suffer from a
plethora of moduli, but these typically also get masses upon supersymmetry
breaking.  Details of these scales depend sensitively on the dynamics; we
will exhibit the mechanism of flux-generated masses in the next section.

There are two broad classes of relevant supersymmetry breaking
mechanisms, \textit{gauge-mediated} and \textit{gravity-mediated}, and
in particular the latter appears to offer the possibility of a large
range of scales.

In gauge-mediated supersymmetry breaking, we imagine that in addition to
the standard model dynamics, the infrared branes produce other dynamics
that breaks supersymmetry and is conveyed to the standard model fields via
a gauge theory messenger.  Such mechanisms have been widely studied; for a
review and references see \cite{Giudice:1998bp}.  It should be noted that
while many of their features are not necessarily modified by virtue of the
warped setting, a TeV-scale gravity scenario does apparently put one strong
constraint on allowed scenarios since the highest allowed scale in the
gauge theory near the SM brane is the TeV scale.  This is problematic in
view of the need for SUSY breaking scales of order 100 TeV to avoid flavor
problems.  We will not explore further aspects of these scenarios in this
paper.

In gravity-mediated scenarios, it appears that there can be a much richer
interplay between the supersymmetry-breaking dynamics and the warping.
For example, first consider supersymmetry breaking produced by gauge
dynamics on other IR branes that are only coupled to standard model
fields via gravity.  In this case, if the hidden-sector supersymmetry
breaking scale is $\Lambda$, we expect that the splittings in the
standard model sector are given by
\begin{equation}
\label{GeneralGravitinoMass}
m_{3/2}\sim  \frac{\Lambda^2}{M_p} \label{mth}
\end{equation}
where here $M_p \sim \bar{M}_D \sim 1$ TeV if the branes are separated on
scales small as compared to the curvature scales/radii of the extra
dimensions, and $M_p \sim M_4$ if the branes are separated on larger
scales.  For $\Lambda\sim$ TeV this can produce the correct splittings if
the effective gravitational mediation scale is $\bar{M}_D$.  However, this
produces splittings that are far too low, ${\cal O}(10^{-4})$ eV, if the
mediation scale goes like $M_4$.

Different scales may.moreover, be generated depending on the
\textit{location} of the supersymmetry breaking in the extra
dimensions; we expect a general relationship
\begin{equation}
\Lambda \sim e^{A_{\rm SUSY}} \Lambda_{SUSY} \label{SUSYscale}
\end{equation}
where $\Lambda_{SUSY}$ is the {\it proper} scale for supersymmetry
breaking (as measured by a higher-dimensional observer in the
supersymmetry breaking region) and $e^{A_{\rm SUSY}}$ is corresponding
warp factor of that region.  For example, one may consider
supersymmetry breaking on some branes that have been raised some
distance up an AdS throat relative to the standard model branes --
although a critical question is how to stabilize such branes.
Alternatively, as mentioned previously, one may generically have
warped compactifications with more than one region with strong warp
factor; standard model branes could be in one region and the
supersymmetry breaking sector in another.  A large relative warp
factor between the two regions can generate a large variation in the
supersymmetry breaking scale.\footnote{Ref.~\cite{Dimopoulos:2001ui}
proposed a different mechanism, {\it tunneling mediation}, for
supersymmetry breaking in such scenarios, although for a large range
of parameters gravity mediation dominates.}  Of course one expects
that the proper scale of supersymmetry breaking is bounded by the
fundamental scale, $\Lambda_{SUSY}\roughly<\bar{M}_D$.  But the relative
factor in (\ref{SUSYscale}) can easily produce a sufficiently large
gravitino mass,
\begin{equation}
m_{3/2} \sim \frac{ e^{2A_{SUSY}} \Lambda_{SUSY}^2}{M_4}\,.
\end{equation}

Indeed, the gravitino mass can also in practice be too \textit{high}.
For example, SUSY breaking in the vicinity of the UV brane could
produce a scale
\begin{equation}
m_{3/2}\sim M_4\,; \label{gravmass}
\end{equation}
we will see a similar phenomenon in models which produce SUSY breaking
through flux in the next section.  However, there is one other
interesting caveat: supersymmetry breaking does not always generate
tree-level masses for superpartners.  This may for example happen if
the K\"ahler and superpotentials of the visible and hidden sectors
completely separate.  Such a mechanism was was proposed in the
``sequestered'' scenario of \cite{Randall:1998uk}.  In this case the
splittings will be produced by loop corrections.  If the gravitino has
a mass given by (\ref{gravmass}), it is effectively removed from the
theory on scales smaller than $M_4$.  One might think this leads to
loop corrections of order $M_4$ to scalar masses in the visible
sector, but note that when one computes the divergent diagrams that
give such masses, the cutoff should actually be the fundamental Planck
scale $\bar{M}_D$.  The important point is that as seen from the
perspective of an observer on the standard model brane, she lives in a
theory that is not supersymmetric, but in which the fundamental scale
and cutoff is $\bar{M}_D\sim$ TeV.  Quantum corrections should thus produce
scalar masses of TeV size.  Similar, though less general,
observations were made in \cite{Luty:2002ff}.

\section{A string theory example: hierarchies from fluxes}
\label{StringSec}

A concrete realization of many of these ideas is provided by the
warped compactification solutions described in \cite{Giddings:2001yu}.
These exhibit some of the basic ideas of the two-brane scenario of
\cite{Randall:1999ee} in a known microscopic theory, namely type IIB
string theory.  They also have other appealing features, as they
improve on a standard phenomenological difficulty of string theory by
stabilizing many of the moduli fields.  Supersymmetry is generically
broken, but both the cosmological constant and, as we discuss below,
masses for ``visible sector'' fields living on a brane are zero at
tree level; this can be related to a ``pseudo-BPS'' condition on the
branes, which we describe shortly.

Specifically, quantized three-form fluxes are introduced inside a
compact six-dimensional manifold, warping a region of the space into
an approximately AdS throat.  The throat is terminated smoothly at
the infrared end by a geometry that is an appropriate analogue of the
Klebanov-Strassler solution \cite{Klebanov:2000hb}, while the unwarped
region of the manifold plays the role of an ultraviolet brane, much as
in \cite{Verlinde:1999fy}.

Mobile branes that fill the non-compact directions are generically
required to be present.  Some of these branes are taken to reside in
the throat region, where the warping induces a hierarchy of scales for
the ``visible sector'' fields on these branes.  In principle one would
like these to be the Standard Model fields, which could perhaps be
realized by placing an additional singularity (or more generally brane
intersections) at the base of the throat, but we will for the moment
content ourselves with the simpler case of the $U(N)$ spectrum of
D3-branes at a generic point.

\subsection{Solutions and geometric scales}

We begin by describing these solutions in more detail.  The bosonic
low-energy action for type IIB supergravity in Einstein frame can be
written (we use the conventions of \cite{Giddings:2001yu}):
\begin{eqnarray}
\label{Action}
S_{IIB}^b = {1 \over 2 \kappa_{10}^2} \int d^{10}x \sqrt{-g} \left\{
{\cal R} - {\partial_M \tau \partial^M \bar\tau \over 2 (\Im \tau)^2}
- { G_{(3)}\cdot {\bar G}_{(3)} \over 12 \, \Im \tau} - {\tilde{F}_{(5)}^2 \over 4
\cdot 5!}\right\} - {1 \over 8i \kappa_{10}^2} \int {C_{(4)} \wedge G_{(3)}
\wedge \overline{G}_{(3)} \over \Im \tau} \label{IIBlag}
\end{eqnarray}
where we have
\begin{eqnarray}
G_{(3)} \equiv F_{(3)} - \tau H_{(3)} \,, &\quad& \tau \equiv C_{(0)}
+ i e^{-\phi} \,, \\ F_{(3)} = dC_{(2)} \,, \quad H_{(3)} = dB_{(2)}
\,, &\quad& \tilde{F}_{(5)} = d C_{(4)} - \df12 C_{(2)} \wedge H_{(3)}
+ \df12 B_{(2)} \wedge F_{(3)} \,.
\nonumber
\end{eqnarray}
Here ${\cal R}$ is the Ricci scalar, $\phi$ is the dilaton, $C_{(0)}$
the RR scalar, $B_{(2)}$ and $C_{(2)}$ the NSNS and RR 2-form
potentials, and $C_{(4)}$ is the RR 4-form potential.  The five-form
field strength $\tilde{F}_{(5)}$ is self-dual,
\begin{eqnarray}
\label{SelfDuality}
\tilde{F}_{(5)} = *\tilde{F}_{(5)} \,,
\end{eqnarray}
which does not follow from the action (\ref{Action}); rather,
(\ref{Action}) is understood to produce the correct equations of
motion when supplemented by (\ref{SelfDuality}).  Dimensionally
reducing the action in a background 5-form field must be done with
care, as we discuss in section~\ref{ModuliSec}. 

It is familiar that a four-dimensional, ${\cal N}=2$ supersymmetric
solution may be obtained from a type II theory by considering a
background geometry of the form $R^4 \times {\cal M}$, where ${\cal
M}$ is a Calabi-Yau threefold.  However, there is a much wider class
of warped compactifications preserving the Poincar\'e symmetry.  The
general Poincar\'e invariant configuration allows the axion-dilaton
scalar to vary over the compact manifold,
\begin{equation}
\tau=\tau(y) \,,
\end{equation}
and allows components of the three- and five-form fluxes in the compact
directions:
\begin{eqnarray}
G_{(3)} &=& {1 \over 3!}\, G_{mnp}(y) \, dy^m  dy^n
dy^p\,, \\
{\tilde F}_{(5)} &=& \partial_m \alpha(y) (1+*)\,  dy^m dx^0 dx^1 dx^2 dx^3
\,.
\label{FiveForm}
\end{eqnarray}
The expression for the five-form is manifestly consistent with
self-duality (\ref{SelfDuality}), and is the most general form
consistent with the Bianchi identity.  Poincar\'e invariance also
allows D3-branes, which will be pointlike in the extra dimensions,
5-branes wrapped on two-cycles, D7-branes wrapped on four-cycles, and
D9-branes.  The metric in general takes the warped form
\begin{eqnarray}
\label{WarpedMetric}
ds^2 = G_{MN} dx^M dx^N = e^{2A(y)} \eta_{\mu\nu} \, dx^\mu dx^\nu +
{g}_{mn}(y) \, dy^m dy^n \,.
\end{eqnarray}
With a typical configuration of branes and fluxes, $g_{mn}$ is no
longer Calabi-Yau. 

Ref.~\cite{Giddings:2001yu} considers a very general class of string
solutions that are obtained by making an additional assumption, and
this class will be the focus of our description for the rest of the
paper.  The assumption is that localized sources such as branes and
orientifold planes must satisfy a BPS-like condition relating their
stress-energy to their D3-brane charge:
\begin{eqnarray}
\label{BPS}
\frac{1}{4}(T_m^m -T_\mu^\mu)^{loc} \geq T_3 \rho_3 \,.
\end{eqnarray}
Here $\rho_3$ is the D3-brane charge density of the localized sources,
and the constant $T_3$ is the D3-brane tension.  This ``pseudo-BPS''
condition roughly states that negative-tension sources (which are of
course allowed in string theory, as for example orientifold planes)
can't be too strong.

Under these added assumptions, \cite{Giddings:2001yu} finds the
general solutions in terms of an underlying Calabi-Yau geometry (or
more generally, in the case with 7-branes, an F-theory background).
The warp factor and five-form are related by
\begin{eqnarray}
e^{4A}=\alpha \,, \label{AlphaWarp}
\end{eqnarray}
the internal metric is {\it conformal} to a Calabi-Yau (or F-theory base)
metric $\tilde{g}$,
\begin{eqnarray}
g_{mn} = e^{-2A} \tilde{g}_{mn} \,,\label{ConfCY}
\end{eqnarray}
the flux must be imaginary self-dual (ISD) in the compact dimensions,
\begin{eqnarray}
 *_6 G_{(3)}= i G_{(3)} \,, \label{ISD}
\end{eqnarray}
where $*_6$ denotes the six dimensional Hodge dual, and finally, the
BPS-like condition (\ref{BPS}) is in fact saturated for all sources.

The presence of localized sources is not an option, but is forced on
us by flux conservation.  Because the $H_{(3)}$ and $F_{(3)}$ fluxes
participate in the 5-form Bianchi identity,
\begin{eqnarray}
\label{Bianchi}
d \tilde{F}_{(5)} = H_{(3)} \wedge F_{(3)} \,,
\end{eqnarray}
together they produce a source of D3-brane charge.  Additional sinks
of D3-brane flux must then be introduced on the compact manifold to
cancel this charge.  Two options were discussed in
ref.~\cite{Giddings:2001yu}: one may quotient the space by a discrete
symmetry so as to introduce orientifold 3-planes, or one may add
7-branes wrapped on four-cycles, both of which carry a D3-brane charge
(in the latter case the charge is induced by the curvature of the
four-cycle).  The 7-branes require a non-Ricci-flat unwarped geometry
as well as a varying axion-dilaton $\tau$, all of which is summarized
as an F-Theory compactification on a Calabi-Yau four-fold $X$.  The
total charge that must vanish is then
\begin{eqnarray}
Q_{D3} = N_{D3} - \tf14 N_{O3} - \frac{\chi(X)}{24} +
\frac{1}{2\kappa_{10}^2 T_3} \int_{\cal M} H_{(3)}\wedge F_{(3)} = 0 \,.
\end{eqnarray}
$\chi(X)$ is the Euler number of $X$, and $N_{D3}$ and $N_{O3}$ denote
the numbers of D3-branes and O3-planes, respectively.  Notice that
with a general choice of fluxes, satisfying this constraint {\it
requires} the presence of some number of explicit D3-branes, on which
gauge dynamics may live.  To avoid the complications of the F-theory
examples, we will often keep the orientifold case in mind, but it
should be remembered that both are possible.

The underlying Calabi-Yau manifold in general has a large collection
of both K\"ahler and complex structure moduli, and this is typically a
problem for string phenomenology.  However, for given quantized
fluxes, the ISD condition (\ref{ISD}) fixes many of these moduli
\cite{Giddings:2001yu}.  This condition can be reexpressed in terms of
the Dolbault cohomology of the CY, as permitting only a primitive
$(2,1)$ form ({\it i.e.}\ a $G_{ij \bar{k}}$ satisfying $g^{j \bar{k}}
G_{ij \bar{k}} =0$) and a $(0,3)$ form.  The former preserves ${\cal
N}=1$ supersymmetry, while the latter breaks all SUSY.  Generically,
one expects both types to be present in a given compact background,
and so SUSY is generally broken.  These models are found classically
to be no-scale models \cite{Cremmer:1983bf}, \cite{Ellis:1983sf}, and
in particular the cosmological constant vanishes despite supersymmetry
breaking.

The 3-form fluxes must satisfy quantization conditions with respect to the
3-cycles on ${\cal M}$; if $C_I$ form a homology basis for three cycles,
\begin{equation}
\int_{C_I} F_{(3)} = (2\pi)^2 \alpha' M_I \,,\  \int_{C_I} 
H_{(3)} = -(2\pi)^2 \alpha' K_I\,. \label{quantcond}
\end{equation}
Consequently they are fixed and do not fluctuate.  A particularly
interesting case, which we will bear in mind as an example, arises if
we work in the vicinity of a conifold point in the Calabi-Yau moduli
space.  Call the degenerating cycle $A$ and its dual cycle $B$, and
suppose we have turned on a flux configuration with
\begin{eqnarray}
\label{KSFluxes}
\int_{A} F_3 =  (2\pi)^2\alpha' M \,,  \quad \quad \int_{B} H_3  =   -(2\pi)^2\alpha' K \,.
\end{eqnarray}
As \cite{Giddings:2001yu} found, this generates an approximately AdS
region, locally resembling the Klebanov-Strassler
geometry \cite{Klebanov:2000hb}.

These particular solutions exemplify the features of warped
compactifications that we have discussed in earlier sections.  The
most fundamental is the warping that arises in the AdS-like region.
The fluxes (\ref{KSFluxes}) produce a relative warp factor
\begin{equation}
e^{A_{min}} \sim \exp(-2\pi K/3Mg_s)\,.  \label{KSwarp}
\end{equation}
between the unwarped region and the bottom of the throat.  

Since the gravitational potential is minimized at the bottom of the
throat, and the configuration is not truly BPS, a reasonable
hypothesis is that a potential for the position of the branes is
generated at loop level and has a minimum when they are at the bottom
of the throat.  
(We will return to related comments when we discuss
generating masses for brane matter.) 
Fields living on branes at the bottom of the throat will perceive a
hierarchy of scales between the apparent $M_4$ and the fundamental
Planck scale $\bar{M}_D$; realistic values of $M_4 \sim 10^{19}$ GeV,
$\bar{M}_D\sim 1$ TeV may be generated through (\ref{mplanck}),
(\ref{KSwarp}) with quite modest values for the flux quanta.  

The rest of the discussion of geometrical scales of section 3 also directly
applies.  If the fundamental Planck scale has been lowered to ${\cal
O}$(TeV), black holes may of course be produced above this threshold on the
SM brane.  Likewise, string states may be produced, at comparable or lower
thresholds depending on the value of the string coupling (thus to agree
with phenomenological bounds, weakly coupled models should instead have
$M_s$ set to ${\cal O}$(TeV) or higher).  Furthermore, the lightest
Kaluza-Klein modes will have masses given by the approximate geometrical
scales at the bottom of the throat; from \cite{Klebanov:2000hb} we find
\begin{eqnarray}
E_{KK}\sim \frac{\bar{M}_s}{g_s M}\,. \label{KKMass}
\end{eqnarray}
This exhausts the discussion of the geometrical scales.  

An important question is to determine the corresponding dynamical
scales, in particular the scale of supersymmetry breaking, the
magnitude of the resulting splittings in supermultiplets in the
visible sector, and the masses of the moduli fields.  We turn to this
task in the coming sections.  We calculate the mass of the gravitino
broken by $(0,3)$ flux as a measure of supersymmetry breaking, as well
as determining the potential for the moduli\footnote{Related work
involving partial SUSY breaking in the unwarped case appeared in
\cite{Curio:2000sc}.}.  We also comment on how supersymmetry breaking is not
communicated to the visible sector fields at tree level, a phenomenon
analogous to the {\em sequestered} scenarios of \cite{Randall:1998uk}.
In the process, we develop expressions for the K\"ahler and
superpotentials for such warped compactifications, which heretofore
have not been calculated with the warping taken into account.


\subsection{The gravitino}

In the absence of $(0,3)$ flux, ${\cal N}=1$ supersymmetry is
preserved in four dimensions.  Correspondingly there is a massless
gravitino.  When SUSY is broken by the flux, the mass $m_{3/2}$ of
this gravitino is a useful measure of the breaking.  We shall begin by
computing this quantity by dimensional reduction of the 10D theory,
and in the process relate this to the expressions for the
superpotential and K\"ahler potential including the effects of
warping.

The equations of motion for the IIB fermions are given in Appendix
B. We find it convenient to work in terms of an action from which
these equations can be derived, and to determine the gravitino mass it
is sufficient to consider the gravitino squared terms:
\begin{eqnarray}
{1 \over \kappa_{10}^2} \int d^{10}x \sqrt{-g} \left\{ i \bar\Psi_M
\Gamma^{MNP} \left( D_N \Psi_P - \tf{i}{2} Q_N \Psi_P - R_P \Psi_N
\right) - \left[ \df{i}{2}\bar\Psi_M \Gamma^{MNP}S_P \Psi^*_N+ {\rm
h.c.}\right] \right\} \,, 
\end{eqnarray}
where $\Psi_M$ is Weyl but not Majorana, $Q_N$ is a composite
connection composed of derivatives of $\tau$ (see \cite{Schwarz:qr}), and the
supercovariantizations are\footnote{The G-field picks up an additional
$\tau$-dependent phase in transforming from the conventions of
\cite{Schwarz:qr}, which we absorb into a redefinition of $\Psi$.}
\begin{eqnarray}
R_M \equiv - {i \over 16 \cdot 5!}
(\Gamma^{M_1 \cdots M_5} \tilde{F}_{M_1 \cdots M_5})
\Gamma_M\,, \quad S_M \equiv {1 \over 96 \, (\Im \tau)^{1/2}} (\Gamma_M^{\;\;
NPQ} G_{NPQ} - 9 \Gamma^{NP} G_{MNP} )\,.
\end{eqnarray}
The supersymmetry variation of the gravitino is
\begin{eqnarray}
\label{GravVar}
\delta \Psi_M = (D_M - \tf{i}{2} Q_M) \varepsilon + R_M \varepsilon +
S_M \varepsilon^* \,,
\end{eqnarray}
where the supersymmetry parameter $\varepsilon$ is a 10D Weyl spinor field.

We must first identify the 4D ${\cal N}=1$ gravitino as a particular
component of the 10D field.  In a warped background satisfying
(\ref{FiveForm}), (\ref{WarpedMetric}), (\ref{AlphaWarp}) without
3-form fluxes, the preserved 4D supersymmetries are associated to
Killing spinors (for more detail, see \cite{Grana:2001xn}\footnote{The
five-form in \cite{Grana:2001xn} is related to our $\tilde{F}_5$ by
$F_{GP} = - \tilde{F}_5$.}):
\begin{eqnarray}
\varepsilon = \zeta(x) \otimes e^{A(y)/2} \chi(y) \,, \quad \quad \tilde{D}_m \chi = 0 \,,
\end{eqnarray}
where we use the tilde to denote the CY metric\footnote{In the case of
an F-theory compactification, $\chi$ is covariantly constant with
respect to $\tilde{D}_m - \tf{i}{2} Q_m$.}. We normalize the covariantly
constant spinor on the unwarped compact space $\chi$ as $\chi^\dagger
\chi = 1$.

Knowing the preserved supersymmetry, we can easily determine the
associated gravitino as the SUSY partner of the 4D graviton.  The
supersymmetry variation of the 4D metric $g_{\mu\nu}$ is
\begin{eqnarray}
\delta g_{\mu\nu} \propto \, \bar\zeta \gamma_\mu \psi_\nu +
 \, \bar\zeta \gamma_\nu \psi_\mu \,,
\end{eqnarray} 
and its 10D counterpart is analogous.  One then finds the 4D gravitino
$\psi_\mu$ embedded in the 10D gravitino as
\begin{eqnarray}
\Psi_\mu = \psi_\mu \otimes e^{A/2} \chi \,.
\end{eqnarray}
It is straightforward to see that under dimensional reduction, the 
Einstein and Rarita-Schwinger terms for the 4D metric $g_{\mu\nu}$ and
gravitino $\psi_\mu$ match the standard form:
\begin{eqnarray}
S = {1 \over \kappa_4^2} \int d^4x \sqrt{-g_4} \left\{ \tf12 {\cal
R}_4 + i \bar\psi_\mu \gamma^{\mu \nu \rho} D_\nu \psi_\rho \right\} \,,
\end{eqnarray}
with the 4D gravitational constant $\kappa_4$ given in terms of the 10
gravitational constant $\kappa_{10}$ and the warped volume $V_w$:
\begin{eqnarray}
{1 \over \kappa_4^2} = {V_w \over \kappa_{10}^2} \,, \quad \quad 
V_w \equiv \int d^6y \sqrt{\tilde{g}_6} \, e^{-4A} \,.
\end{eqnarray}
The 4D gravitino $\psi_\mu$ is massless as long as supersymmetry is
preserved.  The $S_P$ term in the action vanishes, and a possible mass
contribution from the $R_P$ term is canceled by the term in the spin
connection containing a derivative of $A$.

However, in the presence of 3-form fluxes, supersymmetry is generically broken
and the gravitino $\psi_\mu$ acquires a mass.  For a pseudo-BPS
solution, the 5-form/warp factor relation (\ref{AlphaWarp}) persists
and the $R_P$ and spin connection terms continue to cancel.  The mass
term for $\psi_\mu$ is then generated solely from the $S_P$ term in
the 10D action.  Its reduction is straightforward, and one obtains
\begin{eqnarray}
\label{MassTerm}
{1 \over \kappa_{10}^2} \int d^4x \sqrt{-g_4} {1 \over (\Im
\rho)^{3/2}} \left\{ \left( \bar\psi_\mu \gamma^{\mu \nu} \psi_\nu^*
\right) (\tf{i}{48} \int d^6y \sqrt{\tilde{g}_6} {1 \over (\Im
\tau)^{1/2}} \chi^\dagger \tilde{\gamma}^{mnp} \chi^* G_{mnp}) + {\rm
h.c.} \right\} \,,
\end{eqnarray}
where in the above we included the K\"ahler modulus $\rho$ controlling
the overall scale of the compact directions; we discuss $\rho$ in the
next subsection.  This is the proper form for a gravitino mass
term\footnote{The bilinear $\bar\psi_\mu \gamma^{\mu \nu} \psi_\nu^*$
may seem unfamiliar if one is used to 4D gravitini written in Majorana
form, but it is the correct expression for a Weyl gravitino, which
arises naturally from our reduction.}, with
\begin{eqnarray}
\label{GravitinoMass}
m_{3/2} = {1 \over (\Im \rho)^{3/2} (\Im \tau)^{1/2}
 V_w}\left(\tf{1}{24} \int d^6y \sqrt{\tilde{g}_6} \chi^\dagger
 \tilde{\gamma}^{mnp} \chi^* G_{mnp} \right) \,.
\end{eqnarray}
Taking a complex basis $i,j,k,\bar\imath,\bar\jmath,\bar{k}$ for the
Calabi-Yau, we may define the covariantly constant spinor to be the
``lowest weight'' for the Clifford algebra: $\gamma^{\bar\imath} \chi =
0$.  One then sees immediately that only the $(0,3)$ piece of
$G_{(3)}$ contributes to the gravitino mass, as expected.

Given our normalization for $\chi$, we have the relation
\begin{eqnarray}
\chi^\dagger \tilde{\gamma}^{\bar\imath \bar\jmath \bar{k}} \chi^* =
\epsilon^{\bar\imath \bar\jmath \bar{k}} = { \Omega^{\bar\imath
\bar\jmath \bar{k}} \over ||\Omega||} \,,
\end{eqnarray}
up to an undetermined phase, where $\Omega_{ijk}$ is the holomorphic 3-form
of the Calabi-Yau and $3! ||\Omega||^2 = \Omega_{ijk} \overline
\Omega^{ijk}$.  Using
\begin{eqnarray}
||\Omega||^2 V_w &=& ||\Omega||^2 \int d^6y \sqrt{\tilde{g}_6} e^{-4A} \,,
\\ &=& \int e^{-4A} \, \Omega \wedge \overline\Omega \quad \equiv
\quad \omega_w \,,
\nonumber
\end{eqnarray}
we then obtain
\begin{eqnarray}
\label{GravitinoFinalMass}
m_{3/2} = (\Im \rho)^{-3/2} (V_w \omega_w)^{-1/2} (\Im \tau)^{-1/2}
 \left(\tf{1}{4} \int \,\Omega \wedge G \right) \,.
\end{eqnarray} 
In the absence of $G_{(0,3)}$ flux, $m_{3/2} \rightarrow 0$ and 4D
${\cal N}=1$ supersymmetry is restored.  This suggests that the
supersymmetry breaking can be captured in an ${\cal N}=1$ language,
where a gravitino mass can be expressed in terms of the K\"ahler
potential ${\cal K}$ and superpotential $W$ as
\begin{eqnarray}
\label{GravitinoMassPotentials}
m_{3/2} \propto \kappa_4^2 \, e^{{\cal K}/2} \, W \,.
\end{eqnarray}
After discussing the moduli in the next subsection, we will present
values for the K\"ahler and superpotentials, and demonstrate that
(\ref{GravitinoFinalMass}) can be written in the form
(\ref{GravitinoMassPotentials}).  We will estimate the value of
$m_{3/2}$ in subsection~\ref{EstimateSec}.

\subsection{The moduli}
\label{ModuliSec}

Ordinary Calabi-Yau compactifications possess a large number of
moduli, massless fields corresponding to the deformations of the
compact manifold consistent with the Calabi-Yau condition, as well as
the axion-dilaton.  Since the our solutions of have an underlying
Calabi-Yau space, in the absence of fluxes such moduli would also be
present there.  Specifically, the corresponding light fields are the
complex structure moduli $z^\alpha(x)$, the K\"ahler moduli
$\rho^i(x)$, and the axion-dilaton $\tau(x)$.  However, an advantage
of the pseudo-BPS warped compactifications, beyond their original
motivation of solving the hierarchy problem, is that many of the
moduli are fixed by the fluxes, including the dilaton.  This was
understood in \cite{Giddings:2001yu}; one explanation follows from the
assumption of a superpotential of the Gukov-Vafa-Witten form
\cite{Gukov:1999ya},
\begin{eqnarray}
\label{GVW}
W = {a \over \kappa_4^8} \int_{\cal M} \Omega \wedge G \,, 
\end{eqnarray}
(where $a$ is a convention-dependent numerical constant) which is believed
to arise in a wide variety of compactifications of string/M-theory with
fluxes turned on threading calibrated submanifolds.  The flux is fixed, and
the moduli (in this case the complex structure and axion-dilaton) adjust to
minimize F-terms arising from (\ref{GVW}).

In order to give a more complete treatment of these moduli, in this
section we turn to the problem of working out their 4d effective
action and in particular their potential.  The appendix of
\cite{Giddings:2001yu} began the process of explicitly demonstrating
this action, by working out the kinetic terms and potential, together
with their connection with the superpotential (\ref{GVW}), in the
limit where warping can be neglected.  The purpose of the present
section is to give a more complete derivation, and in particular to
find the effective action and K\"ahler and superpotentials in the
presence of non-trivial warping.  This means not just including the
warp factor in the terms studied in \cite{Giddings:2001yu}, but also
incorporating the contributions from the Einstein and five-form terms,
which vanished there.  We proceed by fixing the fluxes---in accord with
the quantization condition (\ref{quantcond})---and investigating the
action for slowly varying fields $z^\alpha(x)$, $\rho^i(x)$, and
$\tau(x)$.

First we calculate the moduli kinetic terms including the warping, and
derive the corresponding warped K\"ahler potential.  The geometrical
moduli fields arise in the metric as
\begin{eqnarray}
ds^2 = e^{2A(y)} e^{-6u(x)} g_{\mu\nu} dx^\mu dx^\nu + e^{-2A(y)}
e^{2u(x)} \left(\tilde{g}_{mn}(y) + T^I(x) \, \delta g_{Imn}(y)\right)
dy^m dy^n \,, \label{moddecomp}
\end{eqnarray}
where the $\delta g_{I}$ are traceless, $\tilde{g}^{mn} \delta g_{Imn} =
0$, so that fluctuations of $e^{2u}$ scale the total volume, while the
$T^I$, which in principle include both the remaining K\"ahler structure
moduli and the complex structure moduli, are volume-preserving at linear
order.  The factor $e^{-6u}$ on the 4D part must be introduced to decouple
$u(x)$ from the 4D graviton.

The kinetic terms for the moduli fields are found by extracting the
quadratic order terms in an expansion of the Einstein-Hilbert term in the 
Lagrangian (\ref{IIBlag}) using the decomposition (\ref{moddecomp}).
These are calculated to be
\begin{eqnarray} 
S_{mod} &=& {1 \over 2 \kappa_{10}^2} \int d^4x \sqrt{-g_4}d^6 y
\sqrt{\tilde{g}_6} \, e^{-4A} \left\{ - {6 \over 4} \, e^{-8u} (\partial_\mu
e^{4u})^2 - {1 \over 4} \partial_\mu T^I \partial^\mu T^J \, \delta
g_{Imn} \delta g_J^{\widetilde{mn}} \right\} \,, \nonumber \\ 
&=& {1 \over  \kappa_4^2}
\int d^4x \sqrt{-g_4} \left\{ - 3 {\partial_\mu \bar\rho \partial^\mu
\rho \over |\rho - \bar\rho|^2} - {1 \over 8 V_w} \, \partial_\mu T^I
\partial^\mu T^J  \int d^6y \sqrt{\tilde{g}_6}\,e^{-4A}  \delta
g_{Imn} \delta g_J^{\widetilde{mn}}  \right\} \,,\label{ModuliKinetic}
\end{eqnarray}
where we have defined the complex field $\rho$ such that $\Im \rho =
e^{4u}$; the real part is a form field that was discussed in
\cite{Giddings:2001yu}.  The moduli space metric for the remaining
fields is seen to be a suitably warped version of the Weil-Petersson
metric.  
From (\ref{Action}), one easily calculates the 4D dilaton kinetic term
to be
\begin{eqnarray}
\label{DilatonKinetic}
S_{dil} = {1 \over \kappa_4^2} \int d^4x \sqrt{-g_4} \left\{ - {\partial_\mu \bar\tau \partial^\mu \tau \over |\tau - \bar\tau|^2} \right\} \,.
\end{eqnarray}
The kinetic terms (\ref{ModuliKinetic}), (\ref{DilatonKinetic}) are
consistent with the K\"ahler potential
\begin{eqnarray}
\label{KahlerPotential}
{\cal K} &=& - 3 \log (-i \left( \rho - \bar{\rho} \right)) -
\log\left(-{ i \over \kappa_4^6} \int d^6y e^{-4A}\sqrt{{\hat g}_6} \,
\right) \nonumber \\ &-& \log \left( -{ i \over \kappa_4^6} \int e^{-4A}
\, \Omega \wedge \overline\Omega \right) - \log (-i \left( \tau -
\bar{\tau} \right)) \,. 
\end{eqnarray}
where the volume piece is computed using the metric
\begin{equation}
{\hat g}(x,y) = \tilde{g}_{mn}(y) + T^I(x) \, \delta g^I_{mn}(y)
\end{equation}
with the overall scale piece removed, as in (\ref{moddecomp}).  Notice
that ${\cal N}=1$ supersymmetry will match the real $T^I$ fields
corresponding to K\"ahler moduli with a set of $p$-form modes into
complex pairs, but these $p$-form fields do not appear in the K\"ahler
potential.

The K\"ahler potential (\ref{KahlerPotential}) has a form quite
similar to that which arises in the unwarped case, with a correction
due to the warp factor inserted to the volume integrals.  The
coefficient of the $\rho$ term identifies this K\"ahler potential as
being of the no-scale form, as noticed in \cite{Giddings:2001yu}.  As
a result the tree-level cosmological constant will vanish despite
supersymmetry breaking.  The result (\ref{KahlerPotential}) is valid
only to leading order in $\alpha'$; some next-to-leading-order results
were examined in \cite{Becker:2002nn} (neglecting warping).  We will
comment more on these corrections later.

A general flux configuration will lift the complex-structure moduli
$z^\alpha$ and fix the dilaton $\tau$.  In order to find this potential,
we assume a general metric that is \textit{constant} in $x$,
\begin{equation}
ds^2 = e^{2A(y)} g_{\mu\nu} dx^\mu dx^\nu + e^{-2A(y)} {\tilde
g}_{mn}(y) \, dy^mdy^n\,; \label{constmet}
\end{equation}
the moduli potential is exhibited from dependence of the action on the
Calabi-Yau metric ${\tilde
g}_{mn}$, as well as the dilaton.  Specifically, 
the effective potential for these is computed from the 
${\cal R}$, $|G_{(3)}|^2$ and $\tilde{F}_{(5)}^2$ terms in the action
(\ref{IIBlag}); these are terms with explicit dependence on the metric in
the compact directions.\footnote{As usual, in the 
F-theory case the dilaton term must be
added to the Ricci term to get the desired result.}

For the metric (\ref{constmet}), the Einstein-Hilbert term can be shown to
give
\begin{equation}
\int d^{10} x {\sqrt -g}{\cal R} = \int d^4 x \sqrt{-g_4} \int d^6 x {\sqrt g_6}
\left[-8(\nabla A)^2 e^{4A} \right]\,. \label{EHcont}
\end{equation}
The action for ${\tilde F}_5$ is more subtle: for a self dual field, it
vanishes. This is part of the usual problem for formulating
an action for self-dual $p$-form field strengths.
One way to obtain a consistent dimensionally-reduced action is to
double the coefficient on the 5-form term, but restrict to components
of $\tilde{F}_{(5)}$ with indices along $R^4$ (or equivalently, restricting it
only to components with no indices along $R^4$).  It is readily checked
that this prescription yields the correct dimensionally reduced equations
of motion for the metric.  Using the expression (\ref{FiveForm}), we find
\begin{equation}
\int d^{10}x \sqrt{-g} \frac{{\tilde F}_5^2}{4\cdot 5!} \rightarrow
\int d^4x \sqrt{-g_4} \int d^6y \sqrt{g_6} \, {e^{-4A} \over 2} \,
(\partial_m\alpha)^2\,.
\end{equation}
Then from 
the relation (\ref{AlphaWarp}), we find a contribution equivalent to
(\ref{EHcont}).  This can be rewritten in terms of the fluxes using the
Bianchi identity (\ref{Bianchi}), which takes the form
\begin{equation}
\square A = {i G_{mnp} *\overline{G}^{mnp} \over 48 \, \Im \tau} + {\rm local} 
\end{equation}
(the localized source terms cancel for sources saturating the pseudo-BPS
condition (\ref{BPS})).
The first term is a total derivative when integrated.  We therefore
find
\begin{equation}
\int d^{10}x \sqrt{-g_{10}}\left[ {\cal R} - \frac{\tilde F_5^2}{4\cdot 5!}\right] =
\int d^4 x \sqrt{-g_4} \int d^6y {\sqrt g_6} \frac{i e^{4A}G_{mnp}*_6 {\bar
G}^{mnp}}{12 \, \Im \tau}\,.
\end{equation}
Combining this with the $G_3$ term then gives
\begin{eqnarray}
\label{GenPot}
S_{\cal V} = {1 \over 2 \kappa_{10}^2} \int d^4x \sqrt{-g_4} \int {e^{4A}
\over 2 \, \Im \tau} \, G_{(3)} \wedge \left(*_6 \overline{G}_{(3)} + i \overline{G}_{(3)} \right)\,.
\end{eqnarray}
Defining imaginary self- and anti-self-dual parts of the flux $G_{(3)}$,
\begin{eqnarray}
\label{SelfDual}
G_{(3)}^\pm = \tf12 (G_{(3)} \pm i *_6 G_{(3)}) \,, \quad \quad *_6
G_{(3)}^\pm = \mp i G_{(3)}^\pm \,,
\end{eqnarray}
we can write the potential (\ref{GenPot}) as 
\begin{eqnarray}
\nonumber
S_{\cal V} &=& {1 \over 2 \kappa_{10}^2} \int d^4x \sqrt{-g_4} \int {i e^{4A} \over \Im \tau} \, G_{(3)} \wedge \overline{G}_{(3)}^+ \,,\\
&=& {1 \over 2 \kappa_{10}^2} \int d^4x \sqrt{-g_4} \int {e^{4A} \over \Im \tau} \, G_{(3)}^+ \wedge *_6 \overline{G}_{(3)}^+ \,,
\label{Potential}
\end{eqnarray}
where in the second line we used the self-duality properties
(\ref{SelfDual}) to relate $\overline{G}_{(3)}^+$ to $*_6
\overline{G}_{(3)}^+$ and to show $G_{(3)}^- \wedge
\overline{G}_{(3)}^+ = 0$.  The potential (\ref{Potential}) has the
form anticipated in \cite{Giddings:2001yu}, 
but with warping included. 

We also anticipate that we should be able to write this potential in
terms of the K\"ahler potential (\ref{KahlerPotential}) and a
superpotential via the usual ${\cal N}=1$ formula
\begin{eqnarray}
\label{PotFromSuperpot}
{\cal V} = \kappa_4^2 \, 
e^{\cal K} \left\{ ({\cal G}^{-1})^{A \bar{B}} D_A W
\overline{D_{\bar{B}} W} - 3 |W|^2 \right\} \, ,
\end{eqnarray}
and it is interesting to check whether the Gukov-Vafa-Witten form (\ref{GVW})
persists in the presence of warping.  
For simplicity we specialize to the case $\tau =$ const. The equation
of motion for $G_{(3)}$ is \cite{Giddings:2001yu}
\begin{eqnarray}
\label{GEOM}
d \Lambda + {i \over \Im \tau} d\tau \wedge {\Re} \Lambda = 0 \,,
\quad \quad \Lambda \equiv e^{4A} *_6 G_{(3)} - i \alpha G_{(3)} \,,
\end{eqnarray}
which then becomes
\begin{eqnarray}
d e^{4A} G_{(3)}^+ = 0 = d *_6 e^{4A} G_{(3)}^+ \,. 
\end{eqnarray}
Consequently $e^{4A} G_{(3)}^+$ is harmonic on the Calabi-Yau, and we
can expand it in a basis of harmonic three-forms.  The analysis now
proceeds analogously to that in \cite{Giddings:2001yu}.  Only the $(3,0)$ and
$(1,2)$ forms have the correct self-duality properties to appear in
the expansion, so we find
\begin{eqnarray}
e^{4A} G_{(3)}^+ = {1 \over \omega_w} \left( \Omega \int G_{(3)} \wedge
\overline\Omega + {\cal G}^{\alpha \bar\beta} \bar\chi_{\bar\beta} \int
G_{(3)} \wedge \chi_\alpha \right) \,,
\end{eqnarray}
where we used $\int G_{(3)}^+ \wedge \overline\Omega = \int G_{(3)}
\wedge \overline\Omega$ and an analogous expression for the basis of
(2,1) forms $\chi_\alpha$, and where ${\cal G}^{\alpha \bar\beta}$ is the
inverse to the metric
\begin{equation}
{\cal G}_{\alpha{\bar \beta}} = - {1 \over \omega_w} \int e^{-4A} \,
\chi_\alpha \wedge {\bar \chi}_{\bar \beta} \,,
\end{equation}
which follows from the K\"ahler potential (\ref{KahlerPotential}).  

Hence the potential is (restoring factors of $\rho$)
\begin{eqnarray}
\nonumber S_{\cal V} &=& {1 \over 2 \kappa_{10}^2} {1 \over \omega_w^2}
\int d^4x \sqrt{-g_4} {1 \over (\Im \rho)^3} \int {e^{-4A} \over \Im \tau}
\, \Biggl[ \Omega \wedge \overline\Omega \int G_{(3)} \wedge
\overline\Omega \int \overline{G}_{(3)} \wedge \Omega \\ && + ({\cal
G}^{-1})^{\alpha \bar\gamma} ({\cal G}^{-1})^{\delta \bar\beta}
\bar\chi_{\bar\beta} \wedge \chi_\alpha \int G_{(3)} \wedge \chi_\delta
\int \overline{G}_{(3)} \wedge \bar\chi_{\bar\gamma} \Biggr]
\label{FinalPotential} \\ &=& {1 \over 2 \kappa_4^2} {1 \over V_w \,
\omega_w} \int d^4x \sqrt{-g_4} {1 \over (\Im \rho)^3} {1 \over \Im \tau}
\left[ \int G_{(3)} \wedge \overline\Omega \int \overline{G}_{(3)} \wedge
\Omega + ({\cal G}^{-1})^{\alpha \bar\beta} \int G \wedge \chi_\alpha \int
\overline{G} \wedge \bar\chi_{\bar\beta} \right] \,, \nonumber
\end{eqnarray}
It is not hard to show that this form can be derived from the K\"ahler
potential (\ref{KahlerPotential}), together with an unwarped GVW
superpotential of the GVW form (\ref{GVW}).  Using the identity
$\partial_\alpha \Omega = k_\alpha \Omega + \chi_\alpha$, where
$k_\alpha$ is a moduli-dependent constant, one may show
\begin{eqnarray}
D_\tau W = - {1 \over (\tau - \bar\tau)} {a \over \kappa_4^8} \int \Omega
\wedge \overline{G} \,, \quad D_\alpha W = { a \over \kappa_4^8} \int
\chi_\alpha \wedge G \,, \quad D_\rho W = - {3 W \over \rho - \bar\rho} \,.
\end{eqnarray}
The potential (\ref{PotFromSuperpot}) may then be
computed.  As in the large-volume case, the $|D_\rho W|^2$ term
cancels $- 3 |W|^2$, producing a no-scale potential.  The other terms
then reproduce (\ref{FinalPotential}), with the overall factors
arising from the K\"ahler potential.

Notice the subtlety of distinguishing $\omega_w$ (which depends on the
complex structure moduli) from $V_w$ (which depends on the K\"ahler
moduli) was essential in making this identification.  In the
large-volume case in \cite{Giddings:2001yu} this subtlety was not
clearly treated.

A check of our derivation of the K\"ahler potential
(\ref{KahlerPotential}) and superpotential (\ref{GVW}) can be obtained
by reproducing the gravitino mass (\ref{GravitinoFinalMass}) from the
formula (\ref{GravitinoMassPotentials}).  We indeed reproduce the
correct form.  This gives us confidence in our results, as well as
reinforcing the ubiquity of he Gukov-Vafa-Witten superpotential.
Although it is generally believed not to receive corrections from the
warp factor, this is to our knowledge the first demonstration that
this is the case.

\subsection{Estimating gravitino and moduli masses}
\label{EstimateSec}

Having obtained an analytic expression for the gravitino and moduli
masses, we would like to estimate their values.  One might intuit that
a bulk field like the gravitino gaining mass from SUSY breaking in the
bulk will have a mass of the same order as the effective scale of
gravity, namely $M_4$.  Indeed this proves generally to be the case.
In calculating bulk quantities it is more convenient to use the
conventions where the warp factor is one at the top of the throat
rather than at the bottom, which we shall do below.

Having succeeded in expressing the gravitino mass
(\ref{GravitinoFinalMass}) in terms of a topological integral
independent of the warping, we can evaluate it in terms of the moduli
of the Calabi-Yau in straightforward fashion.  For the
Klebanov-Strassler fluxes (\ref{KSFluxes}) this was already done in
\cite{Giddings:2001yu}, with the result (using, in our conventions,
$\int_A \Omega = V^{1/2} z$ where $V = \int d^6y \sqrt{\tilde{g}_6}$
is the unwarped volume)
\begin{eqnarray}
\label{EvaluateSuperpot}
\int \Omega \wedge G = ( 2 \pi)^2 (\alpha') V^{1/2} [K \tau z - M {\cal
G}(z)] \,,
\end{eqnarray}
where
${\cal G}(z) = z \log z / (2 \pi i)$ + holomorphic.  Although the
complex structure modulus $z$ is the source of the hierarchy and is
fixed to be exponentially small, the holomorphic part of ${\cal G}(0)$
is generically ${\cal O}(1)$.  Consequently (\ref{EvaluateSuperpot})
is just $(\alpha') V^{1/2}$ times factors of order unity.  Cases with
more general flux configurations will behave similarly: exponentially
small terms in $W(z,\tau)$ will be washed out by ${\cal O}(1)$ terms,
and the overall dimensionful constants will not change.

The expression (\ref{GravitinoFinalMass}) for $m_{3/2}$ also involves
background values of the moduli $\Im \tau$ and $\Im \rho$.  The
axion-dilaton $\Im \tau$ is fixed by the superpotential to be of order
unity\footnote{One needs a slightly more involved set of fluxes than
(\ref{KSFluxes}) to fix the dilaton, see \cite{Giddings:2001yu}.}.  The
volume modulus $\Im \rho$ has a flat potential at tree level.  We have
chosen units, however, where the background value is $\langle \Im \rho
\rangle = 1$; the overall size of the compact manifold is then given
by values for the integrals such as $V$ and $V_w$.  Thus we see that
\begin{eqnarray}
m_{3/2} \sim {(\alpha') V^{1/2} \over V_w} \,.
\end{eqnarray}
When the volume and the warped volume are of the fundamental scale
$M_D \sim M_4$, we find that $m_{3/2} \sim M_4$.

One can estimate the moduli masses in similar fashion.  From
(\ref{PotFromSuperpot}) we read off the form for the moduli potential
\begin{eqnarray}
{\cal V} \sim {1 \over \kappa_4^2} \, m_{3/2}^2 \, {\cal G}^{i \bar{\jmath}}
{ D_i W \overline{D_{\bar{\jmath}} W} \over |W|^2} \,.
\end{eqnarray}
Hence the potential for the moduli is also generically of the scale $M_4$.

The supersymmetry breaking may be heuristically thought of as coming
from the region around the top of the throat.  The $G$-flux vanishes
when the warp factor stops varying, so the source of SUSY breaking is
concentrated in the throat; however it is not localized at the bottom
of the throat, but instead receives its dominant contribution where
the warp factor is largest, which is near the top.  From the point of
view of the earlier discussion on supersymmetry breaking, one may
interpret our result $m_{3/2} \sim M_4$ as
eqs.~(\ref{GeneralGravitinoMass}), (\ref{SUSYscale}) with
$\Lambda_{SUSY} = M_4$ since the breaking is fundamental scale, $M_p =
M_4$ since the SUSY-breaking is well-separated from the visible
sector, and $e^{A_{SUSY}} \sim 1$ since it is near the top of the
throat.

One may be puzzled that the gravitino mass rises so far above the
scale of bulk KK~excitations (\ref{KKMass}).  However, since the
massless graviton stays massless even with the addition of $G$-flux,
the higher excitations of the graviton are protected by 4D general
covariance from receiving mass corrections from the fluxes, and
consequently get mass only from their shape in the compact geometry.
The gravitini have no such protection.

Note that the broken gravitino will also generically receive mixing
terms with the other 7 massive gravitini; for the case we have
outlined all will have masses like $M_4$, the $(0,3)$ flux will be
just as large as the $(2,1)$ flux, and there will not be a region of
energies where ${\cal N}=1$ supersymmetry is a good description.  One
can speculate as to whether one of these other IIB gravitini could
come down in mass as the contribution from the Calabi-Yau
compactification is canceled by the contribution from fluxes (or more
generally, whether an eigenvalue of the gravitino mass matrix might be
particularly small).  Although such a cancellation could conceivably be
engineered at tree level, there is no reason why the mass should
remain small once quantum corrections are included.

All our results hold at leading order in the $\alpha'$-expansion.  It
is likely that $\alpha'$ corrections will destroy the
no-scale structure, giving a potential to the overall volume $\rho$.
A computation of the first subleading order was performed by Becker,
Becker, Haack and Louis \cite{Becker:2002nn}, where a correction to the
K\"ahler potential was found (neglecting warping).  The leading order
correction was not enough to isolate an extremum of the $\rho$
potential, but the corrections involve additional factors of the
superpotential and the volume, which presumably becomes warped. 
The corrections to the potential are of order 
\begin{eqnarray}
\delta {\cal V} \sim { e^{\cal K} |W|^2 \over M_4^2} \sim m_{3/2}^2 M_4^2 \,.
\end{eqnarray}
This suggests that the induced potential for $\rho$ is also of order $M_4$;
whether there is any regime where the no-scale structure is approximately
preserved is not known and would be an important question to answer.

\subsection{Brane matter and Sequestering}

We have estimated the value of the gravitino mass $m_{3/2}$ to be of
order the 4D Planck scale or slightly less.  At first, this seems to
be a phenomenological disaster, since symmetry breaking effects in
visible sector fields might be expected to be as large.  Indeed, it is
easy to see that generic scalar fields $\phi$ with canonical K\"ahler
potential ${\cal K}_\phi \sim \bar\phi \phi$ and no quadratic
contribution to the superpotential receive masses from supersymmetry
breaking on the order of the gravitino mass:
\begin{eqnarray}
D_\phi W \sim W \bar\phi + {\cal O}(\phi^2) \rightarrow {\cal V} \supset e^K |W|^2 \bar\phi \phi \,.
\end{eqnarray}
One might naively believe that brane matter will couple in this
fashion, in which case bulk supersymmetry breaking by ISD fluxes in the
pseudo-BPS spacetimes, despite other nice features, would not be a
viable candidate for phenomenology.

However, one may explicitly calculate the mass induced by the fluxes
for brane fields.  The action for a D-brane is given by the sum of
Born-Infeld and Wess-Zumino actions, given here in string frame,
\begin{eqnarray}
\label{DBIDthree}
S_{D3} = - T_3 \int d^4x e^{-\phi} \sqrt{\det[P( G_{ab} + B_{ab}) +
2 \pi \alpha' F_{ab}]} + \mu_3 \int \sum_i P[C_{(i)}] \wedge e^{2 \pi
\alpha' F - B} \,,  
\end{eqnarray}
where $P$ denotes the pullback of a spacetime quantity to the brane,
$F$ is the worldvolume gauge field and $\mu_3$ and $T_3$ are the
D3-brane charge and tension. In the absence of $G_{(3)}$, the D3-brane
preserves the same supersymmetries as the warped geometry, and thus
there is no potential generated; any potential must appear with the
SUSY breaking.  However, $G_{(3)}$ appears in the D3-brane action
solely through the pullback of the potentials $B_{(2)}$ and $C_{(2)}$.
Since neither potential is polarized along the D3-brane, it is not
hard to convince oneself that all nonvanishing terms in their
pullbacks involve at least one derivative of the brane fields, and
hence cannot generate a potential.  Indeed, one may explicitly check
by examining the three-brane action (\ref{DBIDthree}) that the
relation (\ref{AlphaWarp}) between the warp factor and five-form
guarantees a no-force condition on D3-branes, with gravitational and
RR 5-form potentials canceling.

Furthermore, it was found by Gra\~na \cite{Grana:2002tu} that the
D3-brane fermionic terms do not couple to the imaginary self-dual part
of $G_{(3)}$.  Although other kinds of $G_{(3)}$ flux can lead to
various masses and couplings for brane fermions, the brane is entirely
insensitive to ISD flux.  Consequently, we arrive at the result that
supersymmetry breaking by $(0,3)$ fluxes induces no tree level masses
at all for D3-brane fields.

Vanishing of scalar masses arises from the no-scale structure of the
theory; the additional feature of vanishing fermion masses is
analogous to the sequestered structure proposed in
\cite{Randall:1998uk}, and we shall refer to it as sequestering in
what follows.  The no-scale structure is characterized by the K\"ahler
potential
\begin{eqnarray}
{\cal K} = -3 \log \left[ f_{visible}(X, \bar{X}) +
g_{hidden}(\rho, \bar\rho) \right] \,,
\end{eqnarray}
where $X$ are visible sector  fields and $\rho$ are hidden-sector
fields; supersymmetry breaking in the hidden sector will not be
communicated to the visible sector scalars at tree level.

Ref.~\cite{Randall:1998uk} suggested the naturalness of sequestering
when the visible sector lives on a brane and the SUSY-breaking sector
is physically separated from it in a higher-dimensional space.
However, Anisimov, Dine, Graesser and Thomas (ADGT)
\cite{Anisimov:2001zz, Anisimov:2002az} have pointed out
several examples from string/M-theory, including Type I,
Ho\v{r}rava-Witten, and $Dp$-$Dp'$ systems, where sequestering is not
generic despite the physical separation of sectors on two different
branes.  The reason can be traced to the exchange of bulk
(closed-string) modes at tree level, which can generate contact terms
between the sectors of the order of the gravitino mass.

Our scenario is the first example we know of sequestering in a string
theory background, at least to leading order in the $\alpha'$
expansion.  ADGT \cite{Anisimov:2001zz, Anisimov:2002az} were
aware of the no-scale K\"ahler potential of the pseudo-BPS solutions
of \cite{Giddings:2001yu}, but speculated that even were sequestering
to arise in such models with brane backreaction neglected, such
backreaction would destroy the sequestered form.  This is not the case
for our scenario.  As remarked previously, the pseudo-BPS solutions
can include the presence of certain localized sources---including
D3-branes---in the background.  Hence, although the backreaction of
the D3-branes in the throat will locally change the specific form of
the solution, it will not bring it outside the pseudo-BPS class, and
our conclusions about the lack of tree level masses will persist.

This raises the question as to whether another type of brane known to
sit in the almost-BPS class of objects, such as the 7-brane wrapped on
a 4-cycle, also has worldvolume excitations sequestered from bulk
supersymmetry breaking.  If so, it would provide a richer set of
possibilities for engineering visible sector matter, with the wealth
of possible cycles in the compact space to wrap.  We leave this
question for the future.

In previous sections, we established the warped K\"ahler potential
(\ref{KahlerPotential}) for the bulk moduli.  The D3-brane matter must
enter into the K\"ahler potential as well, and owing to the
sequestered form it must enter in a nontrivial fashion.  A natural
guess is something of the form
\begin{eqnarray}
\label{NewKahler}
{\cal K} = -3 \log \left( -i (\rho - \bar\rho) + K(\bar{X}, X)
\right) + {\cal K}(\tau, \bar\tau) + {\cal K}(z,\bar{z}) \,,
\end{eqnarray}
where $K(\bar{X},X)$ is related to the spacetime K\"ahler
potential for the Calabi-Yau.  This modified K\"ahler potential
preserves the no-scale structure: one may verify that for arbitrary
$K$, the contributions to $|DW|^2$ from $\rho$ and $X$ (including
off-diagonal terms) always combine to give precisely $3 |W|^2$.

The expression (\ref{NewKahler}) also leads to a coupling to the
radial modulus, at leading order, of the form
\begin{eqnarray}
\label{BI}
T_{D3} \int d^4x \sqrt{-g_4} {1 \over \Im \rho } e^{2A}\, \tilde{g}_{i\bar\jmath} \, \partial_\mu X^i \partial^\mu \bar{X}^{\bar\jmath} \,,
\end{eqnarray}
which is the correct power of $\Im \rho$ arising in the BI
action.  The lack of coupling of the dilaton that appears is also
correct for the Einstein-frame action.  We leave further
exploration of this K\"ahler potential, including the coupling of the
complex structure moduli, for future work.

Spartner masses vanish at tree level, but as discussed in section 4, should
receive corrections at loop level.  From the point of view of an observer
on the brane, supersymmetry is broken  -- the gravitino is eliminated from the
low energy spectrum.  Thus generic loop corrections are expected to raise
mass scales to the cutoff scale.  However, as we have emphasized, for an
observer on an IR brane in a TeV-scale gravity scenario, the fundamental
scale is ${\cal O}$(TeV), and this is where the cutoff on loop momenta
should be placed: above this scale, one encounters strongly-coupled
gravitational physics.  Thus spartner masses are generically expected to be
around a TeV in such a scenario, which is a reasonable answer for
phenomenology.  

Note that part of the original motivation for the sequestered
scenarios of \cite{Randall:1998uk} was to have a situation where the
dominant contribution to spartner masses was through anomaly mediation
\cite{Randall:1998uk, Giudice:1998xp}, with mass scale
\begin{eqnarray}
m_{AMSB} \sim b_0 \left( g^2 \over 16 \pi^2 \right) m_{3/2} \,, 
\end{eqnarray}
where $b_0$ is the one-loop beta function coefficient (for the scalars
there is an an additional constant from the anomalous dimensions).
Given that in the present case the gravitino mass is far above the
effective cutoff scale of ${\cal O}$(TeV), it seems that this formula
cannot give the correct masses here; rather it appears that the masses
arise from generic loop corrections.  A better understanding of the
role of AMSB in this model could be of interest.

Another possible way to exploit the no-scale structure is to find a
background in which the (0,3) component of $G_{(3)}$ can be switched
off, and to break SUSY on another set of branes situated in the middle
of the throat.  The K\"ahler potential becomes
\begin{eqnarray}
{\cal K} = -3 \log \left( -i (\rho - \bar\rho) + f(\bar{X}, X) +
g(\bar{Y} , Y) \right) + {\cal K}(\tau, \bar\tau) + {\cal
K}(z,\bar{z}) \,,
\end{eqnarray}
where $Y$ are the hidden sector fields.  Again, as far as SUSY
breaking is concerned this has the no-scale form.  The location of the
hidden sector brane could be tuned to provide the right amount of SUSY
breaking; this is a fine-tuning, but it preserves the other advantage
of AMSB, that it addresses the supersymmetric flavor problem.  This
sort of ``brane'' SUSY breaking is much more prevalent in the
literature than the ``bulk'' SUSY breaking we have examined for much
of this paper.

\subsection{Summary of phenomenology}
\label{SummarySec}

Since this section has been rather long and technical, we give an overview
of its essential results here.  

String theory solutions found in \cite{Giddings:2001yu} provide a
non-trivial example of many of the warped compactification ideas discussed
in the first four sections, and in particular can be arranged to generate a
hierarchy through warping and thus produce a TeV-scale gravity scenario.
This means that geometrical scales will be realized as was discussed
in sections \ref{WarpedSec} and \ref{GeomSec}.  In particular, for an observer
on the IR brane where we imagine standard model physics residing, the
fundamental Planck scale will be reached at scattering energies ${\cal
O}$(TeV), and we can envision string and black hole production taking
place at such energies.  Kaluza-Klein masses are even lighter, and are
given in terms of the flux quanta by eq.~(\ref{KKMass}).

These solutions arise by considering close analogs of Calabi-Yau manifolds
with three-form fluxes frozen into their geometry.  These fluxes break
supersymmetry.  They also generate a potential for many of the moduli
fields that would otherwise be massless in a standard Calabi-Yau
compactification.  The gravitino mass is given in
eq.~(\ref{GravitinoFinalMass}), and can be estimated to be of order the
four-dimensional Planck scale, $10^{19}$ GeV.  The moduli kinetic terms are
given in eq.~(\ref{ModuliKinetic}), and the potential for moduli in
eq.~(\ref{Potential}).  This lifts the complex structure moduli and the
dilaton to have masses also generically of order $10^{19}$ GeV.  The action
for the moduli, and for the gravitino, can be conveniently summarized in
supergravity language in terms of a K\"ahler potential,
eq.~(\ref{KahlerPotential}), and a superpotential, given by
eq.~(\ref{GVW}).  These explicitly include the effects of the warping.

Although supersymmetry is broken at a high scale, at tree level the
cosmological constant vanishes and matter fields on an IR brane have
vanishing masses.  For scalars, this statement corresponds to the fact
that we are dealing with a {\it no-scale} model.  This structure also
extends to fermion matter, resulting in a {\it sequestered} structure.
This structure survives brane back-reaction.  Spartners are however
expected to get masses from loop corrections, but since the
fundamental scale for brane matter, and hence the relevant cutoff,
lies at the TeV scale, these masses are expected to be TeV-size.

\section{Conclusions}
\label{ConclusionSec}

We have discussed a number of generic features of the scales and
thresholds in warped compactifications, and illustrated them in the
special case of the solutions of IIB string theory given in
\cite{Giddings:2001yu}.  The latter solutions in particular offer
possible solutions to some of the difficult problems of string
phenomenology.  Supersymmetry is broken by three-form fluxes frozen
into the geometry, and a potential for a large number of otherwise
problematic moduli is generated at the same time.  Spartner masses are
not generated at tree level, but in such a TeV-scale gravity scenario
are expected to receive loop corrections of TeV magnitude.

While these certainly seem like interesting successes, it should be
borne in mind that there are a number of other problems that must be
resolved in order to find solutions of string theory that realize
TeV-scale gravity and reproduce a realistic phenomenology including
the standard model.  (Several of these are also problems also for more
traditional Planck-scale compactifications of string theory, so do not
discriminate against TeV-scale scenarios.)  One obvious question is
how to realize the structure of the standard model within the general
framework of this kind of solution.  Many ideas have occurred in the
literature, involving intersecting branes and branes at singularities,
and it may be possible to combine these scenarios with a framework
like that presented here, but clearly there is some non-trivial work
to be done; some interesting recent progress in this direction
includes \cite{Cremades:2002dh, Blumenhagen:2002wn}.  Particularly
challenging issues include reproducing the gauge groups and matter
representations, with reasonable couplings, of the standard model;
addressing baryon and lepton number violation, and reproducing the
relation between the gauge coupling constants that can otherwise be
taken to indicate matching via renormalization group running to a
grand unified scale.  A second problem is that of the remaining
moduli; in particular, K\"ahler moduli are not stabilized by the
fluxes we consider, and thus must be fixed by another mechanism.  This
is a generic problem, since the overall scale of the compact manifold
is generically a K\"ahler modulus.  (For another approach, see
\cite{Silverstein:2001xn}.)  Corrections at higher order in string
loops or $\alpha'$ (see {\it e.g.}  \cite{Becker:2002nn}) may play a
role, but it is difficult to see they do so {\it and} maintain
reasonable mass scales.  In particular, we must ultimately face the
thorny problem of the cosmological constant, which here as in other
scenarios with broken supersymmetry would appear to take a value that
is far too large.

\bigskip
\bigskip
\centerline{\bf Acknowledgments}
\medskip
The authors would like to thank N. Arkani-Hamed, T. Banks, M. Dine,
S. Kachru, R. Kallosh, J. Polchinski, R. Schoen, and S. Thomas for
very valuable conversations.  In addition, SBG would like to thank the
Newton Institute, SLAC, and in particular the Stanford ITP for their
hospitality while much of this work was done.  The research of OD was
supported by the National Science Foundation under grant PHY99-07949.
The research of SBG was supported in part by Department of Energy
under contract DE-FG-03-91ER40618, and by the David and Lucile Packard
foundation.

\section*{Appendix A: Conventions}

We work in mostly-plus signature in both ten and four dimensions.  We
use $M,N$ for 10D indices, $\mu,\nu$ for 4D indices, and $m,n$ for
generic 6D indices; the latter can in turn be divided into holomorphic
$i,j$ and antiholomorphic $\bar{\imath}, \bar{\jmath}$ indices with respect to
the complex structure of the Calabi-Yau threefold.

Ten dimensional gamma-matrices $\Gamma^M$ are $32 \times 32$ matrices.
They decompose into a product of $4 \times 4$ 4D matrices $\gamma^\mu$
and $8 \times 8$ 6D matrices $\tilde\gamma^i$ as follows:
\begin{eqnarray}
\Gamma^\mu = e^{-A} \gamma^\mu \otimes I \,, \quad \quad \Gamma^m =
e^A \gamma_5 \otimes \tilde\gamma^m \,,
\end{eqnarray}
where
\begin{eqnarray}
\{ \gamma^\mu , \gamma^\nu \} = 2 g^{\mu\nu} \,, \quad \quad
\{ \tilde\gamma^m , \tilde\gamma^n \} = 2 \tilde{g}^{mn} \,.
\end{eqnarray}
The chirality matrices are related as $\Gamma_{11} = \gamma_5
\tilde{\gamma}_{\cal M}$, and obey ${\gamma_5}^2 = {\tilde\gamma_{\cal M}}^2
= {\Gamma_{11}}^2 = 1$.

The ten-dimensional gravitino $\Psi_M$ is Weyl (but not Majorana):
\begin{eqnarray}
\Gamma_{11} \Psi_M = - \Psi_M \,.
\end{eqnarray}
It decomposes into the 4D gravitino as $\Psi_\mu = \psi_\mu \otimes
e^{A/2} \chi$, for which we have for our class of solution
\begin{eqnarray}
\gamma_5 \psi_\mu = \psi_\mu \, \quad \quad \tilde\gamma_{\cal M} \chi = - \chi \,.
\end{eqnarray}

\section*{Appendix B: Fermionic equations of motion for IIB supergravity}

The fermionic equations of motion for type IIB supergravity (to linear
order in fermions) are presented in equations (4.6), (4.12) of
\cite{Schwarz:qr} with coefficients given by (4.8), (4.14).  However,
the derivative in these equations contains supercovariantizations
involving three- and five-form fluxes not explicitly recorded there.
The complete definition is however implicit in other expressions given
in \cite{Schwarz:qr}, notably the supersymmetric variations of the
fermionic equations of motion (4.7), (4.10), (4.13) and (4.15), and
can be deduced from these.  We collect the complete equations here for
future convenience.

In this appendix we use the conventions of \cite{Schwarz:qr} for the
fields, although we use our index conventions. We indicate how to pass
to our field conventions at the end.

The dilatino equation of motion (4.6) of \cite{Schwarz:qr} is
\begin{eqnarray}
\Gamma^M \hat{D}_M \lambda = {i \kappa \over 240} \Gamma^{M_1 \ldots
M_5} \lambda F_{M_1 \ldots M_5} + {\cal O}(\Psi^3) \,,
\end{eqnarray}
where the supercovariant derivative of the dilatino is
\begin{eqnarray}
\label{DilSuper}
\hat{D}_M \lambda &=& D_M \lambda  - \kappa T
\Psi_M - \kappa U \Psi^*_M\,, \\ T &=& - {i \over 24} \Gamma^{MNP}
G_{MNP}\,, \quad U = {i \over \kappa} \Gamma^M P_M \, .
\end{eqnarray}
Here $D_M= \nabla_M - \tf{i}{2} Q_M$ contains the ordinary covariant
derivative including the spin connection $\nabla_M$, and a composite
connection $Q_M$ composed of the complex scalar, while $P_M$ is the
field strength for the complex scalar.  The gravitino equation of
motion is
\begin{eqnarray}
\Gamma^{MNP} \hat{D}_N \Psi_P = - {i \over 2} \Gamma^P \Gamma^M
\lambda^* P_P - {i \kappa \over 48} \Gamma^{NPQ}\Gamma^M \lambda G^*_{NPQ}
+ {\cal O}(\Psi^3) \,,
\end{eqnarray}
where the supercovariant derivative acting on the gravitino is
\begin{eqnarray}
\label{GravSuper}
\hat{D}_N \Psi_P &=& D_N \Psi_P  - \kappa R_P \Psi_N - \kappa S_P \Psi_N^*\,, \\
R_M &=& {i \over 480} (\Gamma^{M_1 \cdots
M_5} F_{M_1 \cdots M_5}) \Gamma_M\,, \quad 
S_M = {1 \over 96} (\Gamma_M^{\;\; NPQ} G_{NPQ} - 9 \Gamma^{NP} G_{MNP} )\,.
\end{eqnarray}
A supercovariant derivative in a general supergravity theory consists of
the ordinary covariant derivative supplemented with terms involving the
gravitino such that the supersymmetry variation of the combined terms does
not contain any derivatives of the supersymmetry parameter $\varepsilon$.
These expressions arise naturally in supergravity equations of motion, as
the variation of a one-derivative fermionic equation must be a bosonic
equation with two derivatives on the fields, and hence a derivative may not
be spared to act on $\varepsilon$.  Equations (\ref{DilSuper}),
(\ref{GravSuper}) constitute the general form for supercovariantization of
the derivative in an arbitrary supergravity theory with fermionic
supersymmetry variations
\begin{eqnarray}
\delta \Psi_M &=& {1 \over \kappa} D_M \varepsilon + R_M \varepsilon + S_M \varepsilon^* \,, \\
\delta \lambda &=& T \varepsilon + U \varepsilon^* \,.
\end{eqnarray}
In our conventions, Schwarz's constant $\kappa=1$, and should not be
confused with our $\kappa_{10}$ which is an overall coefficient in the
action and does not appear in the equations of motion.  The relation
between Schwarz's $F$ and $G$ and the $\tilde{F}_5$ and $G_3$ of this
paper is
\begin{eqnarray}
F_{Sch} = - {1 \over 4} \tilde{F}_5 \,, \quad \quad G_{Sch} = {i e^{i \theta} \over \sqrt{\Im \tau}} G_3 \,, \quad e^{i\theta} \equiv \left( {1 + i \bar\tau \over 1 - i \tau} \right)^{1/2} \,.
\end{eqnarray}
For relations involving the complex scalar, see \cite{Grana:2001xn};
note that \cite{Grana:2001xn} use an $F = 4 F_{sch}$, and
consequently for them $\alpha = - e^{4A}$.



\begingroup\raggedright\endgroup


\begin{thebibliography}{10}

\bibitem{Arkani-Hamed:1998rs}
N.~Arkani-Hamed, S.~Dimopoulos and G.~R.~Dvali,
``The hierarchy problem and new dimensions at a millimeter,''
Phys.\ Lett.\ B {\bf 429}, 263 (1998)
[arXiv:hep-ph/9803315].

\bibitem{Randall:1999ee}L.~Randall and R.~Sundrum,``A large mass hierarchy from a small extra dimension,''Phys.\ Rev.\ Lett.\  {\bf 83}, 3370 (1999)[arXiv:hep-ph/9905221].


\bibitem{Antoniadis:1998ig}

I.~Antoniadis, N.~Arkani-Hamed, S.~Dimopoulos and G.~R.~Dvali ``New
dimensions at a millimeter to a Fermi and superstrings at a TeV,'' Phys.\
Lett.\ B {\bf 436}, 257 (1998) [arXiv:hep-ph/9804398].


\bibitem{Giddings:2001yu}
S.~B.~Giddings, S.~Kachru and J.~Polchinski,
``Hierarchies from fluxes in string compactifications,''
arXiv:hep-th/0105097, to appear in Phys. Rev. D.

\bibitem{Verlinde:1999fy}
H.~Verlinde,
``Holography and compactification,''
Nucl.\ Phys.\ B {\bf 580}, 264 (2000)
[arXiv:hep-th/9906182].

\bibitem{Chan:2000ms}
C.~S.~Chan, P.~L.~Paul and H.~Verlinde,
``A note on warped string compactification,''
Nucl.\ Phys.\ B {\bf 581}, 156 (2000)
[arXiv:hep-th/0003236].

\bibitem{Klebanov:2000hb}
I.~R.~Klebanov and M.~J.~Strassler,
``Supergravity and a confining gauge theory: Duality cascades and  
chiSB-resolution of naked singularities,''
JHEP {\bf 0008}, 052 (2000)
[arXiv:hep-th/0007191].

\bibitem{Dasgupta:1999ss}
K.~Dasgupta, G.~Rajesh and S.~Sethi, ``M theory, orientifolds and G-flux,"
JHEP {\bf 08} 023 (1999)
[arXiv:hep-th/9908088].

\bibitem{Gukov:1999ya}
S.~Gukov, C.~Vafa and E.~Witten,
``CFT's from Calabi-Yau four-folds,''
Nucl.\ Phys.\ B {\bf 584}, 69 (2000)
[Erratum-ibid.\ B {\bf 608}, 477 (2000)]
[arXiv:hep-th/9906070].

\bibitem{Grana:2002tu}
M.~Grana,
``D3-brane action in a supergravity background: The fermionic story,''
arXiv:hep-th/0202118.

\bibitem{Randall:1998uk}
L.~Randall and R.~Sundrum,
``Out of this world supersymmetry breaking,''
Nucl.\ Phys.\ B {\bf 557}, 79 (1999)
[arXiv:hep-th/9810155].


\bibitem{Cremmer:1979up}
E.~Cremmer and B.~Julia,
``The SO(8) Supergravity,''
Nucl.\ Phys.\ B {\bf 159}, 141 (1979).

\bibitem{deWit:1984va}
B.~de Wit and H.~Nicolai,
``A New SO(7) Invariant Solution Of D = 11 Supergravity,''
Phys.\ Lett.\ B {\bf 148}, 60 (1984).

\bibitem{vanNieuwenhuizen:ri}
P.~van Nieuwenhuizen and N.~P.~Warner,
``New Compactifications Of Ten-Dimensional And Eleven-Dimensional Supergravity On Manifolds Which Are Not Direct Products,''
Commun.\ Math.\ Phys.\  {\bf 99}, 141 (1985).

\bibitem{Duff:hr}
M.~J.~Duff, B.~E.~Nilsson and C.~N.~Pope,
``Kaluza-Klein Supergravity,''
Phys.\ Rept.\  {\bf 130}, 1 (1986).

\bibitem{Abe:2001wn}
T.~Abe {\it et al.}  [American Linear Collider Working Group Collaboration],
``Linear collider physics resource book for Snowmass 2001,''
in {\it Proc. of the APS/DPF/DPB Summer Study on the Future of Particle Physics (Snowmass 2001) } ed. R.~Davidson and C.~Quigg,
SLAC-R-570
{\it Resource book for Snowmass 2001, 30 Jun - 21 Jul 2001, Snowmass, Colorado}.


\bibitem{Giddings:2000ay}
S.~B.~Giddings and E.~Katz, ``Effective
theories and black hole production in warped compactifications,''
J.\ Math.\ Phys.\ {\bf 42}, 3082 (2001)
[arXiv:hep-th/0009176].



\bibitem{Giddings:2001bu}
S.~B.~Giddings and S.~Thomas,
``High energy colliders as black hole factories: The end of short  distance physics,''
arXiv:hep-ph/0106219.

\bibitem{Dimopoulos:2001hw}
S.~Dimopoulos and G.~Landsberg,
``Black holes at the LHC,''
Phys.\ Rev.\ Lett.\  {\bf 87}, 161602 (2001)
[arXiv:hep-ph/0106295].

\bibitem{Thorne:ji}
K.~S.~Thorne,
``Nonspherical Gravitational Collapse: A Short Review,''
{\it  In *J R Klauder, Magic Without Magic*, San Francisco 1972, 231-258}.

\bibitem{Penrose} R.~Penrose, unpublished; reported in \cite{D'Eath:hb}.

\bibitem{D'Eath:hb}
P.~D.~D'Eath and P.~N.~Payne,
``Gravitational Radiation In High Speed Black Hole Collisions. 1. Perturbation Treatment Of The Axisymmetric Speed Of Light Collision,''
Phys.\ Rev.\ D {\bf 46}, 658 (1992).

\bibitem{D'Eath:hd}
P.~D.~D'Eath and P.~N.~Payne,
``Gravitational Radiation In High Speed Black Hole Collisions. 2. Reduction To Two Independent Variables And Calculation Of The Second Order
News Function,''
Phys.\ Rev.\ D {\bf 46}, 675 (1992).

\bibitem{D'Eath:qu}
P.~D.~D'Eath and P.~N.~Payne,
``Gravitational Radiation In High Speed Black Hole Collisions. 3. Results And Conclusions,''
Phys.\ Rev.\ D {\bf 46}, 694 (1992).

\bibitem{Banks:1999gd}
T.~Banks and W.~Fischler,
``A model for high energy scattering in quantum gravity,''
arXiv:hep-th/9906038.

\bibitem{Argyres:1998qn} P.~C.~Argyres, S.~Dimopoulos and
J.~March-Russell, ``Black holes and sub-millimeter
dimensions,''
Phys.\ Lett.\ B {\bf 441}, 96 (1998)
[arXiv:hep-th/9808138].
9808138;
\bibitem{Emparan:2000rs}
R.~Emparan, G.~T.~Horowitz and
R.~C.~Myers,
``Black holes radiate mainly on the brane,''
Phys.\
Rev.\ Lett.\ {\bf 85}, 499 (2000)
[arXiv:hep-th/0003118].
= HEP-TH 0003118;

\bibitem{Eardley:2002re}
D.~M.~Eardley and S.~B.~Giddings,
``Classical black hole production in high-energy collisions,''
arXiv:gr-qc/0201034.

\bibitem{Giddings:2001ih}
S.~B.~Giddings,
``Black hole production in TeV-scale gravity, and the future of high  energy physics,''
in {\it Proc. of the APS/DPF/DPB Summer Study on the Future of Particle Physics (Snowmass 2001) } ed. R.~Davidson and C.~Quigg,
arXiv:hep-ph/0110127.

\bibitem{Giddings:2002av}
S.~B.~Giddings,
``Black holes at accelerators,''
arXiv:hep-th/0205027.

\bibitem{Giddings:2002kt}
S.~B.~Giddings,
``Black holes in the lab?,''
arXiv:hep-th/0205205.


\bibitem{Horowitz:1996nw}
G.~T.~Horowitz and J.~Polchinski,
``A correspondence principle for black holes and strings,''
Phys.\ Rev.\ D {\bf 55}, 6189 (1997)
[arXiv:hep-th/9612146].


\bibitem{Bowick:1989us}M.~J.~Bowick and S.~B.~Giddings,``High Temperature Strings,''Nucl.\ Phys.\ B {\bf 325}, 631 (1989).

\bibitem{Dimopoulos:2001qe}S.~Dimopoulos and R.~Emparan,``String balls at the LHC and beyond,''Phys.\ Lett.\ B {\bf 526}, 393 (2002)[arXiv:hep-ph/0108060].

\bibitem{Cullen:2000ef}S.~Cullen, M.~Perelstein and M.~E.~Peskin,
``TeV strings and collider probes of large extra dimensions,''Phys.\ Rev.\ D {\bf 62}, 055012 (2000)[arXiv:hep-ph/0001166].

\bibitem{Giudice:1998bp}
G.~F.~Giudice and R.~Rattazzi,
``Theories with gauge-mediated supersymmetry breaking,''
Phys.\ Rept.\  {\bf 322}, 419 (1999)
[arXiv:hep-ph/9801271].


\bibitem{Dimopoulos:2001ui}S.~Dimopoulos, S.~Kachru, N.~Kaloper, A.~E.~Lawrence and E.~Silverstein,``Small numbers from tunneling between brane throats,''Phys.\ Rev.\ D {\bf 64}, 121702 (2001)[arXiv:hep-th/0104239].

\bibitem{Luty:2002ff} M.~A.~Luty, ``Weak scale supersymmetry without
weak scale supergravity,''arXiv:hep-th/0205077.


\bibitem{Cremmer:1983bf} E.~Cremmer, S.~Ferrara, C.~Kounnas and
D.~V.~Nanopoulos, ``Naturally Vanishing Cosmological Constant In N=1
Supergravity,'' Phys.\ Lett.\ B {\bf 133}, 61 (1983).

\bibitem{Ellis:1983sf}
J.~R.~Ellis, A.~B.~Lahanas, D.~V.~Nanopoulos and K.~Tamvakis,
``No - Scale Supersymmetric Standard Model,''
Phys.\ Lett.\ B {\bf 134}, 429 (1984).

\bibitem{Curio:2000sc} G.~Curio, A.~Klemm, D.~Lust and S.~Theisen,
``On the vacuum structure of type II string compactifications on
Calabi-Yau spaces with H-fluxes,'' Nucl.\ Phys.\ B {\bf 609}, 3 (2001)
[arXiv:hep-th/0012213].

\bibitem{Schwarz:qr} J.~H.~Schwarz, ``Covariant Field Equations Of
Chiral N=2 D = 10 Supergravity,'' Nucl.\ Phys.\ B {\bf 226}, 269
(1983).

\bibitem{Grana:2001xn}
M.~Grana and J.~Polchinski, ``Gauge / gravity duals with holomorphic dilaton,''
Phys.\ Rev.\ D {\bf 65}, 126005 (2002)
[arXiv:hep-th/0106014].

\bibitem{Becker:2002nn}
K.~Becker, M.~Becker, M.~Haack and J.~Louis,
``Supersymmetry breaking and alpha' corrections to flux induced  potentials,''
JHEP {\bf 0206}, 060 (2002)
[arXiv:hep-th/0204254].


\bibitem{Anisimov:2001zz}
A.~Anisimov, M.~Dine, M.~Graesser and S.~Thomas,
``Brane world SUSY breaking,''
Phys.\ Rev.\ D {\bf 65}, 105011 (2002)
[arXiv:hep-th/0111235].

\bibitem{Anisimov:2002az}
A.~Anisimov, M.~Dine, M.~Graesser and S.~Thomas,
``Brane world SUSY breaking from string/M theory,''
arXiv:hep-th/0201256.

\bibitem{Giudice:1998xp}
G.~F.~Giudice, M.~A.~Luty, H.~Murayama and R.~Rattazzi,
``Gaugino mass without singlets,''
JHEP {\bf 9812}, 027 (1998)
[arXiv:hep-ph/9810442].


\bibitem{Cremades:2002dh}
D.~Cremades, L.~E.~Ibanez and F.~Marchesano,
``Standard model at intersecting D5-branes: Lowering the string scale,''
arXiv:hep-th/0205074.

\bibitem{Blumenhagen:2002wn}
R.~Blumenhagen, V.~Braun, B.~Kors and D.~Lust,
``Orientifolds of K3 and Calabi-Yau manifolds with intersecting D-branes,''
JHEP {\bf 0207}, 026 (2002)
[arXiv:hep-th/0206038].

\bibitem{Silverstein:2001xn}
E.~Silverstein,
``(A)dS backgrounds from asymmetric orientifolds,''
arXiv:hep-th/0106209.
\end{thebibliography}
\end{document}